\documentclass[aps, prd, letterpaper, 12pt, nofootinbib, superscriptaddress, longbibliography, notitlepage, floatfix, preprint]{revtex4-2}

\usepackage[utf8]{inputenc}
\usepackage{amsmath,amssymb,amsfonts}
\usepackage{mathrsfs}
\usepackage[table]{xcolor}
\usepackage{booktabs}
\usepackage{graphicx}
\usepackage{subfigure}
\usepackage{slashed}
\usepackage{array}

\definecolor{hgreen}{rgb}{0,.7,0}
\definecolor{hred}{rgb}{.7,0,0}
\definecolor{hblue}{rgb}{0,0,.7}

\usepackage[colorlinks=true,
linkcolor=hblue,
citecolor=hgreen,
filecolor=hblue,
urlcolor=hred]{hyperref}

\allowdisplaybreaks

\begin{document}

\preprint{MITP-26-013}

\title{\texorpdfstring{\LARGE\color{black!70} \bf\boldmath\textsl{Large Hadronic Effects in} $B \to K^*\!\mu\mu$?}{Large Hadronic Effects in B to Kstar mu mu ?}}

\author{Wolfgang~Altmannshofer}
\email{waltmann@ucsc.edu}
\affiliation{Department of Physics, University of California, Santa Cruz, and Santa Cruz Institute for Particle Physics, Santa Cruz, CA 95064, USA}

\author{Samuel~G.~Christensen}
\email{sgchrist@ucsc.edu}
\affiliation{Department of Physics, University of California, Santa Cruz, and Santa Cruz Institute for Particle Physics, Santa Cruz, CA 95064, USA}

\author{Peter~Stangl}
\email{peter.stangl@uni-mainz.de}
\affiliation{Institute of Physics, Johannes Gutenberg University Mainz, Staudingerweg 7, 55128 Mainz, Germany}

\begin{abstract}
Recent results from LHCb have confirmed the long-standing $P_5^\prime$ anomaly, an intriguing discrepancy in the angular distribution of the $B \to K^* \mu^+\mu^-$ decay that might be a sign of new physics. In addition, the new results hint at a non-zero value for $S_7$, another observable that characterizes the $B \to K^* \mu^+\mu^-$ angular distribution. We stress that a non-zero $S_7$ cannot be explained by heavy new physics but instead necessarily requires a sizable hadronic effect that introduces a strong phase. We argue that, under plausible assumptions, the hadronic effect is of the correct size to also explain $P_5^\prime$. The direct CP asymmetry in $B \to K^* \mu^+\mu^-$ emerges in principle as a clean probe of new physics in such a scenario. We show that a combined fit of hadronic parameters and Wilson coefficients retains sensitivity to new physics and we find strong bounds on imaginary parts of new physics Wilson coefficients.
\end{abstract}

\maketitle

\newpage
\tableofcontents

\section{Introduction} \label{sec:intro}

For over a decade, rare flavor-changing neutral current (FCNC) decays of $B$ mesons have provided some of the most intriguing hints for physics beyond the Standard Model (SM). Among these, the decay $B \to K^* \mu^+ \mu^-$ has been of particular interest due to persistent deviations from SM predictions in angular observables~\cite{LHCb:2013ghj, LHCb:2015svh, ATLAS:2018gqc, LHCb:2020lmf, LHCb:2020gog, CMS:2024atz, LHCb:2025mqb}. The most prominent of these is the $P_5^\prime$ anomaly, which has been consistently observed by both LHCb~\cite{LHCb:2020lmf} and CMS~\cite{CMS:2024atz}. In addition, measurements of the $B \to K^* \mu^+ \mu^-$ branching ratio and the branching ratios of the related modes $B \to K \mu^+ \mu^-$ and $B_s \to \phi \mu^+ \mu^-$ are significantly below the SM predictions~\cite{LHCb:2014cxe, LHCb:2021zwz, LHCb:2025mqb, CMS-PAS-BPH-23-003}. Recently, the LHCb collaboration presented an updated analysis of the $B \to K^* \mu^+ \mu^-$ decay, utilizing the full Run 1 and Run 2 data set, corresponding to an integrated luminosity of $8.4\,\text{fb}^{-1}$~\cite{LHCb:2025mqb}. The update not only confirms the anomalous behavior in $P_5^\prime$, but also reveals a tantalizing new pattern: a consistent, albeit statistically modest ($\sim 1\sigma - 1.5\sigma$ per $q^2$ bin), downward shift in the angular observable $S_7$ across most of the $q^2$ region below the $J/\psi$ resonance.

\begin{figure}[tb]
\centering
\includegraphics[width=0.48\textwidth]{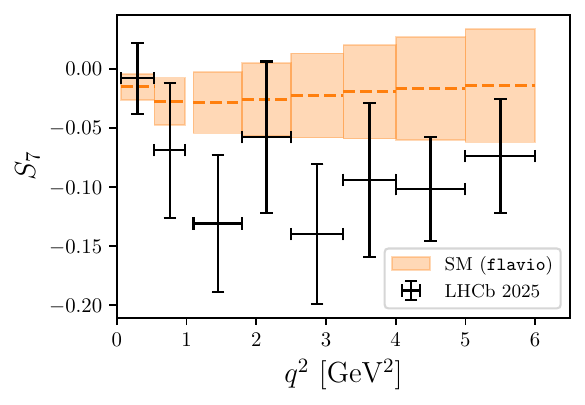} \quad
\includegraphics[width=0.48\textwidth]{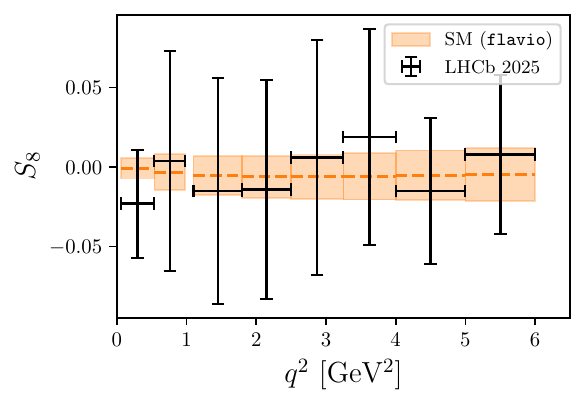} \\[8pt]
\includegraphics[width=0.48\textwidth]{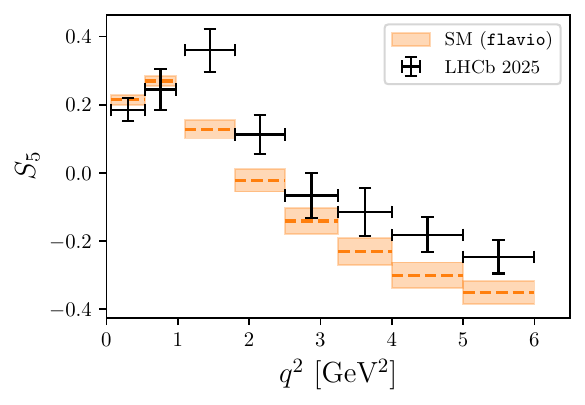} \quad
\includegraphics[width=0.48\textwidth]{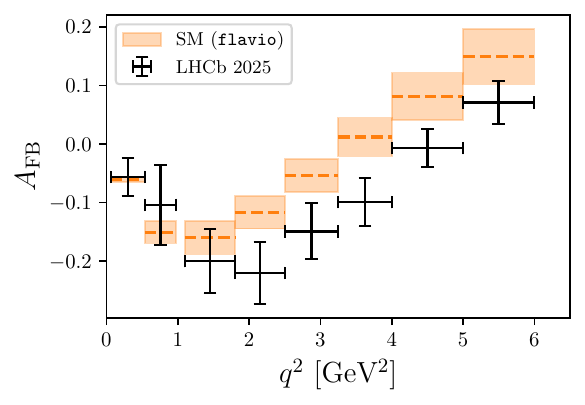} \\[8pt]
\includegraphics[width=0.48\textwidth]{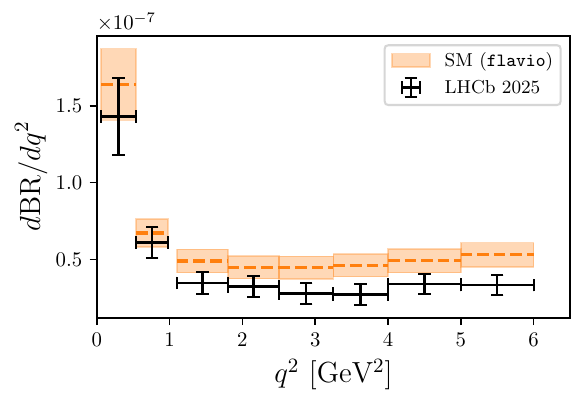}
\caption{The $B \to K^* \mu^+ \mu^-$ angular observables $S_7$ (top left), $S_8$ (top right), $S_5$ (center left), and $A_\text{FB}$ (center right), as well as the differential branching ratio $d\text{BR}/dq^2$ (bottom) as a function of $q^2$. We compare the finely binned experimental results from LHCb~\cite{LHCb:2025mqb} (black) to the SM predictions from \texttt{flavio} (orange) in the $q^2$ region below the $J/\psi$ resonance.}
\label{fig:intro}
\end{figure}

For convenience, the situation is summarized in figure~\ref{fig:intro}.
The upper-left plot shows the finely binned LHCb measurement of $S_7$ comparing it to the SM prediction. Here and in the following, ``SM prediction'' refers to the SM prediction as implemented in \texttt{flavio}~\cite{Straub:2018kue}, including its default model of non-local hadronic effects (see Section~\ref{sec:numerics} for details). The hadronic model is not expected to give reliable results close to the $J/\psi$ resonance and we thus do not show SM predictions above a di-muon invariant mass squared of $q^2 = 6$\,GeV$^2$. In the upper-right plot, we show the related observable $S_8$ that is in good agreement with the SM predictions and compatible with zero. Instead of $P_5^\prime$~\cite{Descotes-Genon:2012isb, Descotes-Genon:2013vna}, we show in the center-left plot the observable $S_5$~\cite{Altmannshofer:2008dz} which is proportional to $P_5^\prime$ and thus shows the same anomaly. Compared to the SM prediction, the experimental data shows a consistent upwards shift. The center-right plot shows the CP-averaged forward-backward asymmetry $A_\text{FB}$, another ``anomalous'' angular observable. The experimental measurement is consistently below the SM prediction. Finally, in addition to the angular observables, also the measured $B \to K^* \mu^+ \mu^-$ branching ratio is not in good agreement with the SM predictions. As shown in the bottom plot, it is consistently low in the low-$q^2$ range.

While deviations in $S_5$, $A_\text{FB}$ and the branching ratio can be interpreted as new physics modifications of the short-distance Wilson coefficients, in particular $C_9$ (see for example~\cite{Altmannshofer:2021qrr, Ciuchini:2021smi, Gubernari:2022hxn, Greljo:2022jac, Ciuchini:2022wbq, Alguero:2023jeh, Wen:2023pfq, Altmannshofer:2023uci, Guadagnoli:2023ddc, Hurth:2023jwr, Bordone:2024hui, Hurth:2025vfx}), the observable $S_7$ is fundamentally different. $S_7$ is the counterpart of the $T$-odd CP asymmetry $A_7$~\cite{Bobeth:2008ij} and as such it is proportional to a strong phase. The observation of a sizable $S_7$ would thus indicate the presence of a sizable strong phase. Such strong phases, however, are naively expected to be very small in the low-$q^2$ region. In fact, in the heavy quark limit, the $B \to K^* \mu^+ \mu^-$ decay amplitudes factorize, and strong phases only arise at next-to-leading order in the strong coupling $\alpha_s$ or at higher orders in the $\Lambda_{\text{QCD}}/m_b$ power expansion. Furthermore, in the kinematic region below the open-charm threshold ($q^2 < 4m_c^2$) the perturbative charm-loop contribution is purely real. A sizable $S_7$ thus points towards non-perturbative, long-distance hadronic effects. Crucially, this line of argument remains robust even in the presence of heavy new physics, which can introduce new weak phases but no new strong phases. Consequently, a non-zero $S_7$ cannot be accommodated by heavy new physics operators but requires a substantial long-distance hadronic effect.

In this work, we argue that the emerging hint of a non-zero $S_7$ could provide the missing key to understanding the $B \to K^* \mu^+ \mu^-$ anomalies. If the deviation in $S_7$ is not a statistical fluctuation, it establishes the presence of sizable, complex hadronic contributions. We argue that the imaginary parts of these hadronic effects required to explain $S_7$ are of the same size as real parts that explain the $B \to K^* \mu^+ \mu^-$ angular anomalies in $S_5$ and $A_\text{FB}$ and the low branching ratio. Furthermore, we explore the phenomenological consequences of the sizable strong phase, showing that it can in principle enhance the new physics sensitivity of the direct CP asymmetry $A_\text{CP}$ and other CP asymmetries.

Several recent works have explored the interplay between hadronic effects and short-distance contributions in $B \to K^* \mu^+\mu^-$ observables using different strategies. Among those most closely related to our study are: Ref.~\cite{Ciuchini:2022wbq} which performed a global analysis of $b \to s \ell \ell$ decays including simultaneous fits of hadronic parameters and Wilson coefficients, focusing on real Wilson coefficients; Ref.~\cite{Bordone:2024hui}, where the $b \to s \ell \ell$ data were interpreted in terms of an effective $q^2$ dependent and helicity dependent coefficient $C_9(q^2,\lambda)$ (see~\cite{Altmannshofer:2015sma, Descotes-Genon:2015uva, Altmannshofer:2017fio} for the first studies of such type); and Ref.~\cite{Hurth:2025vfx} which considered fits including the latest LHCb data but varies either hadronic parameters or Wilson coefficients, but not both simultaneously. In our study we perform simultaneous fits of hadronic parameters and complex Wilson coefficients, emphasizing the role of $S_7$ in determining imaginary parts of the hadronic parameters and the sensitivity of our fits to imaginary parts of the Wilson coefficients.

The remainder of this paper is organized as follows. In Section~\ref{sec:theory}, we introduce the theoretical framework, including the effective Hamiltonian and our phenomenological model of the non-local hadronic matrix elements. In Section~\ref{sec:analytical}, we present analytic approximations for the angular observables in the heavy quark limit in order to illustrate the role of strong phases and their impact on the observables. In Section~\ref{sec:numerics}, we confront our framework with the available experimental data. We first perform fits of the hadronic parameters to the LHCb measurements of $B \to K^* \mu^+\mu^-$ observables and then explore the interplay between hadronic effects and possible new physics contributions through combined fits including short-distance Wilson coefficients. We conclude in Section~\ref{sec:conclusion} with a summary of our results and an outlook.

\section{Theoretical Framework} \label{sec:theory}

In this section, we outline the theoretical framework to describe the $B \to K^* \mu^+ \mu^-$ decay. In Section~\ref{sec:heff} we define the effective Hamiltonian and the relevant operators, while in Section~\ref{sec:hadronic}, we introduce a simple parameterization of the hadronic effects, which will be useful to obtain the analytic expressions in Section~\ref{sec:analytical}.

\subsection{Effective Hamiltonian and Helicity Amplitudes} \label{sec:heff}

To establish the theoretical framework, we begin by reviewing the standard effective Hamiltonian for $b \to s \ell^+ \ell^-$ transitions. We parameterize new physics contributions by Wilson coefficients of dimension-6 operators:
\begin{equation}
  \mathcal H_\text{eff}
  = \mathcal H_\text{eff}^\text{SM} - \frac{4G_F}{\sqrt{2}} V_{tb}V_{ts}^* \frac{e^2}{16\pi^2}
  \sum_{i=7,9,10}
  \left(\Delta C_i \mathcal{O}_i + C^{\prime}_i \mathcal{O}^{\prime}_i \right)
  + \text{h.c.}\,.
\end{equation}
In this  work, we focus on the electromagnetic dipole operator $\mathcal{O}_7$ and the semileptonic vector and axial-vector operators $\mathcal{O}_9$ and $\mathcal{O}_{10}$, alongside their chirality-flipped counterparts:
\begin{align}
\mathcal{O}_7 &= \frac{m_b}{e} (\bar{s} \sigma_{\alpha\beta} P_{R} b) F^{\alpha\beta}\,,
&
\mathcal{O}_7^{\prime} &= \frac{m_b}{e} (\bar{s} \sigma_{\alpha\beta} P_{L} b) F^{\alpha\beta}\,,\label{eq:O7}
\\
\mathcal{O}_9 &= (\bar{s} \gamma_{\alpha} P_{L} b)(\bar{\mu} \gamma^\alpha \mu)\,,
&
\mathcal{O}_9^{\prime} &= (\bar{s} \gamma_{\alpha} P_{R} b)(\bar{\mu} \gamma^\alpha \mu)\,,\label{eq:O9}
\\
\mathcal{O}_{10} &= (\bar{s} \gamma_{\alpha} P_{L} b)( \bar{\mu} \gamma^\alpha \gamma_5 \mu)\,,
&
\mathcal{O}_{10}^{\prime} &= (\bar{s} \gamma_{\alpha} P_{R} b)( \bar{\mu} \gamma^\alpha \gamma_5 \mu)\,.\label{eq:O10}
\end{align}
The Wilson coefficients of the operators $\mathcal{O}_{7}$, $\mathcal{O}_{9}$, and $\mathcal{O}_{10}$ consist of the SM contribution and a potential new physics correction $C_i = C_i^\text{SM} + \Delta C_i$. The primed coefficient $C_i^\prime$ are to a good approximation entirely due to new physics. In addition to $\mathcal{O}_{7}$, $\mathcal{O}_{9}$, and $\mathcal{O}_{10}$, the effective Hamiltonian in the SM, $\mathcal H_\text{eff}^\text{SM}$, also contains the chromo-dipole operator $\mathcal O_8$ and several four-quark operators (see e.g.~\cite{Chetyrkin:1996vx, Bobeth:1999mk, Altmannshofer:2008dz})

The $B \to K^* \mu^+ \mu^-$ decay amplitude is obtained by evaluating the hadronic matrix elements of the above operators between the initial $B$ and final $K^*$ states. The corresponding helicity amplitudes (with the $K^*$ meson helicity $\lambda \in \{+, -, 0\}$) serve as convenient building blocks for the $B \to K^* \mu^+\mu^-$ observables. The most relevant amplitudes are (see e.g.~\cite{Jager:2012uw})
\begin{align}
H_V^{\lambda} &\propto C_9 \tilde V_{L\,\lambda} + C_9^\prime \tilde V_{R\, \lambda} + \frac{2 m_b m_B}{q^2} \Big( C_7 \tilde T_{L\,\lambda} + C_7^\prime \tilde T_{R\,\lambda} \Big) - \frac{m_B^2}{q^2} 16\pi^2 h_\lambda(q^2) \,, \\
H_A^{\lambda} &\propto C_{10} \tilde V_{L\,\lambda} + C_{10}^\prime \tilde V_{R\, \lambda} \,,
\end{align}
where $\tilde V_{L\,\lambda}$, $\tilde V_{R\,\lambda}$, $\tilde T_{L\,\lambda}$, and $\tilde T_{R\,\lambda}$ are $q^2$-dependent helicity form factors defined in~\cite{Jager:2012uw}. All non-local effects, stemming for example from charm-loops, are parameterized by functions of $q^2$ that are denoted as $h_\lambda(q^2)$.

\subsection{Parameterization of Hadronic Effects} \label{sec:hadronic}

The dominant source of theoretical uncertainty in $B \to K^* \mu^+ \mu^-$ arises from the non-local matrix elements. Known effects include for example perturbative loop contributions from four-quark operators (often absorbed into $q^2$-dependent effective Wilson coefficients $C_7^\text{eff}(q^2)$ and $C_9^\text{eff}(q^2)$), spectator scattering, and weak annihilation contributions. We incorporate additional ``sub-leading'' effects without relying on theoretical estimates by adopting a simple empirical parameterization.\footnote{See~\cite{Jager:2012uw, Lyon:2014hpa, Jager:2014rwa, Bobeth:2017vxj, Blake:2017fyh, Chrzaszcz:2018yza, Ciuchini:2021smi, Ciuchini:2022wbq, Gubernari:2022hxn, LHCb:2023gel, LHCb:2023gpo, LHCb:2024onj, Bordone:2024hui, Isidori:2024lng, Isidori:2025dkp, Hurth:2025vfx, Capdevila:2025drq} for other approaches to the non-local effects.} Adopting the approach that is outlined in~\cite{Altmannshofer:2014rta, Altmannshofer:2021qrr} and implemented in \texttt{flavio}, we include the subleading non-local hadronic contributions as shifts of $h_\lambda(q^2)$ that we expand in powers of $q^2$ in the following way
\begin{eqnarray} \label{eq:h0}
\Delta h_0(q^2) &=& - \frac{1}{16\pi^2} \left(  a_0 + b_0 \,\frac{q^2}{m_B^2} + \dots \right) \tilde T_{L \, 0} ~, \\ \label{eq:hminus}
\Delta h_-(q^2) &=& -  \frac{1}{16\pi^2} \left(  a_- + b_- \,\frac{q^2}{m_B^2} + \dots \right) \tilde T_{L \, -}  ~, \\ \label{eq:hplus}
\Delta h_+(q^2) &=&  -  \frac{1}{16\pi^2} \left(  a_+ + b_+ \frac{q^2}{m_B^2} + \dots \right) \tilde T_{R \, +} ~,
\end{eqnarray}
where the ellipses correspond to higher powers in $q^2$.
We treat the constants $a_{0,\pm}$ and $b_{0,\pm}$ as free complex parameters.
Note that the corrections to the $\lambda = +$ helicity amplitude are expected to be higher order in the $1/m_b$ power expansion~\cite{Beneke:2001at, Jager:2012uw}, which suggests $a_+, b_+ \ll a_- , b_-$. We do not impose this hierarchy in our parameterization, but will explore in Section~\ref{sec:numerics} if the data is compatible with this expectation.

To build an intuitive understanding of the impact of these hadronic contributions, it is instructive to analyze approximate analytical expressions for the amplitudes and the key angular observables. In particular, we will work in the heavy quark limit which allows us to express the helicity form factors in terms of the two functions $\xi_\perp$ and $\xi_\parallel$ which we parameterize as in~\cite{Beneke:2001at}
\begin{equation} \label{eq:ff}
\xi_\perp(q^2) = \frac{\xi_\perp(0)}{(1 - q^2/m_B^2)^2}~,\quad \xi_\parallel(q^2) = \frac{\xi_\parallel(0)}{(1 - q^2/m_B^2)^3}~.
\end{equation}
To first approximation, this gives the following helicity amplitudes
\begin{eqnarray} \label{eq:HVplus}
 H_V^{+} &\propto& \left( \frac{m_B^2}{q^2} \left( \frac{2m_b}{m_B} C_7^\prime + a_+ \right) + C_9^\prime + b_+ \right) \frac{\xi_\perp(0)}{1 - q^2/m_B^2} ~, \\
 H_A^{+} &\propto& C_{10}^\prime \frac{\xi_\perp(0)}{1 - q^2/m_B^2} ~, \\  \label{eq:HVminus}
 H_V^{-} &\propto& -\left(\frac{m_B^2}{q^2} \left( \frac{2m_b}{m_B} C_7 + a_- \right) + C_9 + b_- \right) \frac{\xi_\perp(0)}{1 - q^2/m_B^2} ~, \\
 H_A^{-} &\propto& -C_{10} \frac{\xi_\perp(0)}{1 - q^2/m_B^2} ~, \\  \label{eq:HV0}
 H_V^{0} &\propto& - \frac{m_B^2}{4 m_{K^*} \sqrt{q^2}} \left( \frac{2m_b}{m_B} \big(C_7 - C_7^\prime \big) + C_9 - C_9^\prime + a_0 + b_0 \frac{q^2}{m_B^2}\right) \frac{\xi_\parallel(0)}{1 - q^2/m_B^2} ~, \\
 H_A^{0} &\propto& - \frac{m_B^2}{4 m_{K^*} \sqrt{q^2}} \left(C_{10} - C_{10}^\prime \right) \frac{\xi_\parallel(0)}{1 - q^2/m_B^2} ~.
\end{eqnarray}
In the $H_V^{+}$ amplitude, the real parts of the hadronic parameters $a_+$ and $b_+$ effectively mimic shifts in the real parts of the Wilson coefficients $C_7^\prime$ and $C_9^\prime$, respectively. Similarly, in the $H_V^{-}$ amplitude, the real parts of $a_-$ and $b_-$ mimic shifts in the real parts of $C_7$ and $C_9$.
For the longitudinal amplitude $H_V^{0}$, the real part of $a_0$ induces effects that are degenerate with shifts in a specific linear combination of $C_7^{(\prime)}$ and $C_9^{(\prime)}$, while the parameter $b_0$ introduces a qualitatively distinct $q^2$ dependence. Additional $q^2$ dependence in the hadronic contributions may arise from higher-order terms in the parameterizations~\eqref{eq:h0},~\eqref{eq:hminus}, and~\eqref{eq:hplus}.
The mentioned degeneracies between hadronic parameters and Wilson coefficients apply only to their real parts. The imaginary parts behave differently: going from the $\bar B \to \bar K^*$ decay to the CP-conjugate $B \to K^*$ decay, the imaginary parts of the Wilson coefficients change sign, whereas the imaginary parts of the hadronic parameters remain unchanged.

\section{Analytic Approximations for Observables} \label{sec:analytical}

In this section, we utilize the approximate helicity amplitudes in the heavy quark limit that incorporate our simple hadronic model to derive transparent approximate formulas for various $B \to K^* \mu^+ \mu^-$ observables in terms of the Wilson coefficients and the hadronic parameters $a_\lambda$, $b_\lambda$ and confront them with the data.
We consider $S_7$ and $S_8$ (Section~\ref{sec:S7S8}), $A_\text{FB}$ and $S_5$ (Section~\ref{sec:AFBS5}), the angular observables $S_3$ and $S_9$ (Section~\ref{sec:RH}), and the direct CP asymmetry $A_\text{CP}$ (Section~\ref{sec:ACP}).
The discussion in this section is largely qualitative, with a more accurate numerical study following in Section~\ref{sec:numerics}.

\subsection{The Angular Observables \texorpdfstring{$S_7$}{S7} and \texorpdfstring{$S_8$}{S8}} \label{sec:S7S8}

The CP-averaged angular observables $S_7$ and $S_8$~\cite{Altmannshofer:2008dz} share many similarities. They are the CP conserving counterparts of the CP asymmetries $A_7$ and $A_8$~\cite{Bobeth:2008ij} which are $T$-odd. The $T$-oddness of $A_7$ and $A_8$ implies that both $S_7$ and $S_8$ vanish in the absence of strong phases.
Neglecting finite muon mass effects, the SM predictions for the observables $S_7$ and $S_8$ can be approximated by
\begin{multline} \label{eq:S7_approx}
S_7^\text{SM}(q^2) \simeq -\frac{4 m_{K^*} \sqrt{q^2}}{m_B^2} \frac{\xi_\perp(0)}{\xi_\parallel(0)} C_{10}^\text{SM} \bigg(\frac{m_B^2}{q^2} \Big( \mathrm{Im}\, a_-  - \mathrm{Im}\,a_+ \Big) \\ + \mathrm{Im}\,b_-  - \mathrm{Im}\,b_+ - \mathrm{Im}\,a_0 - \mathrm{Im}\,b_0 \frac{q^2}{m_B^2} \bigg) \frac{1}{D^\text{SM}} ~,
\end{multline}
\begin{multline} \label{eq:S8_approx}
S_8^\text{SM}(q^2) \simeq \frac{2 m_{K^*} \sqrt{q^2}}{m_B^2} \frac{\xi_\perp(0)}{\xi_\parallel(0)} \Bigg[ \left(\frac{2m_b}{m_B} C_7^\text{SM} + C_9^\text{SM} + \mathrm{Re}\,a_0 + \mathrm{Re}\,b_0 \frac{q^2}{m_B^2}  \right) \\
\times \bigg(\frac{m_B^2}{q^2} \Big( \mathrm{Im}\,a_- + \mathrm{Im}\,a_+ \Big) + \mathrm{Im}\,b_- + \mathrm{Im}\,b_+ \bigg) \\
-  \left( C_9^\text{SM} + \mathrm{Re}\,b_- + \mathrm{Re}\,b_+ + \frac{ m_B^2}{q^2} \left( \frac{2m_b}{m_B} C_7^\text{SM} + \mathrm{Re}\,a_- + \mathrm{Re}\,a_+ \right) \right) \\ \times  \left(\mathrm{Im}\,a_0 + \mathrm{Im}\,b_0 \frac{q^2}{m_B^2} \right)  \Bigg] \frac{1}{D_\text{SM}}~.
\end{multline}
with the denominator factor
\begin{equation} \label{eq:DSM}
 D_\text{SM} = \big(C_{10}^\text{SM}\big)^2 + \left(\frac{2m_b}{m_B} C_7^\text{SM} + C_9^\text{SM} + \mathrm{Re}\,a_0 + \mathrm{Re}\,b_0 \frac{q^2}{m_B^2}  \right)^2 +  \left( \mathrm{Im}\,a_0 + \mathrm{Im}\,b_0 \frac{q^2}{m_B^2} \right)^2~.
\end{equation}
The factor $D_\text{SM}$ reflects the dependence of the CP-averaged branching ratio $d\text{BR}/dq^2$ on the Wilson coefficients and hadronic parameters.

The above expressions transparently show that $S_7$ and $S_8$ can only be non-zero in the presence of strong phases, which in the used parameterization are provided by the imaginary parts of the hadronic parameters $a_\lambda$ and $b_\lambda$. This conclusion continues to hold for arbitrary new physics contributions to the Wilson coefficients. The corresponding approximate expressions that generalize~\eqref{eq:S7_approx} and~\eqref{eq:S8_approx} in the presence of new physics are given in appendix~\ref{app:NP}.

Linearizing the SM predictions in the hadronic parameters $a_\lambda$ and $b_\lambda$, we find
\begin{multline}
S_7^\text{SM}(q^2) \simeq 1.46 \Big( \mathrm{Im}\,a_- - \mathrm{Im}\,a_+ \Big) \frac{\text{GeV}}{\sqrt{q^2}} + 0.053 \Big( \mathrm{Im}\,b_- - \mathrm{Im}\,b_+ - \mathrm{Im}\,a_0 \Big) \frac{\sqrt{q^2}}{\text{GeV}}  \\+ 0.0019~ \mathrm{Im}\,b_0 \frac{q^2 \sqrt{q^2}}{\text{GeV}^3}   ~,
\end{multline}
\begin{multline}
S_8^\text{SM}(q^2) \simeq \bigg( 0.66 \Big( \mathrm{Im}\,a_- + \mathrm{Im}\,a_+ \Big) + 0.086 ~\mathrm{Im}\,a_0 \bigg) \frac{\text{GeV}}{\sqrt{q^2}} + \bigg(0.024 \Big( \mathrm{Im}\,b_- + \mathrm{Im}\,b_+ \Big) \\ - 0.027 ~\mathrm{Im}\,a_0 + 0.0031 ~\mathrm{Im}\,b_0 \bigg) \frac{\sqrt{q^2}}{\text{GeV}} - 0.00097~\mathrm{Im}\,b_0 \frac{q^2 \sqrt{q^2}}{\text{GeV}^3}  ~,
\end{multline}
where we used $m_B \simeq 5.28$\,GeV, $m_b \simeq 4.18$\,GeV, $m_{K^*} \simeq 0.89$\,GeV, $\xi_\perp(0)/\xi_\parallel(0) \simeq 3.1$~\cite{Jager:2012uw}, $C_9^\text{SM} \simeq 4.2$, $C_{10}^\text{SM} \simeq -4.1$, and $C_7^\text{SM} \simeq -0.3$.

Interestingly, the latest LHCb results~\cite{LHCb:2025mqb} indicate preference for a non-zero $S_7$, while the observable $S_8$ is compatible with zero. Our setup contains a sufficient number of free parameters to accommodate such a scenario.
In particular, $S_7 \simeq -0.1$ to $-0.15$ across most of the low $q^2$ range (see figure~\ref{fig:intro}) na{\"i}vely suggests $\mathrm{Im}\,a_- \sim \mathcal O(0.1)$ and $\mathrm{Im}\,b_-$, $\mathrm{Im}\,a_0 \sim \mathcal O(1)$.
Also $\mathrm{Im}\,a_+$ and $\mathrm{Im}\,b_+$ may play a role, but they are expected to be smaller based on heavy quark expansion arguments. An approximately vanishing $S_8$ appears accidental is such a scenario and requires some amount of cancellation between different imaginary parts.

While the expressions~\eqref{eq:S7_approx} and~\eqref{eq:S8_approx} are not expected to be very accurate approximations, they still indicate the generic size of hadronic parameters that is required to explain the data on $S_7$ and $S_8$. Intriguingly, imaginary parts $\mathrm{Im}\,b_-, \mathrm{Im}\,a_0 \sim \mathcal O(1)$ appear to be a plausible origin of the non-zero $S_7$. In Section~\ref{sec:numerics}, we will test these order-of-magnitude expectations against fits of the hadronic parameters to data.

\subsection{The Angular Anomalies: \texorpdfstring{$A_\text{FB}$}{AFB} and \texorpdfstring{$S_5$}{S5}} \label{sec:AFBS5}

Next, we investigate simple approximations for the CP-averaged forward-backward asymmetry $A_\text{FB}$ and the angular observable $S_5$. Following the same steps as for $S_7$ and $S_8$, we arrive at the SM predictions
\begin{multline}
A_\text{FB}^\text{SM}(q^2) \simeq -\frac{24 m_{K^*}^2 q^2}{m_B^4}  \frac{\xi^2_\perp(0)}{\xi^2_\parallel(0)} C_{10}^\text{SM} \left( C_9^\text{SM} + \mathrm{Re}\,b_- + \frac{m_B^2}{q^2} \left( \frac{2m_b}{m_B} C_7^\text{SM} + \mathrm{Re}\,a_- \right) \right) \frac{1}{D_\text{SM}} ~,
\end{multline}
\begin{multline}
S_5^\text{SM}(q^2) \simeq \frac{4 m_{K^*} \sqrt{q^2}}{m_B^2} \frac{\xi_\perp(0)}{\xi_\parallel(0)} C_{10}^\text{SM} \bigg( \frac{m_B^2}{q^2} \left( \frac{2 m_b}{m_B}C_7^\text{SM} + \mathrm{Re}\,a_- + \mathrm{Re}\,a_+ \right)  \\
+ \frac{2m_b}{m_B} C_7 + 2 C_9^\text{SM} + \mathrm{Re}\,b_- + \mathrm{Re}\,b_+ + \mathrm{Re}\,a_0 + \mathrm{Re}\,b_0 \frac{q^2}{m_B^2} \bigg) \frac{1}{D_\text{SM}}~.
\end{multline}
with $D_\text{SM}$ already defined in equation~\eqref{eq:DSM}.

Linearizing the SM predictions in the $a_\lambda$ and $b_\lambda$, we find the following shifts in the observables due to the hadronic parameters
\begin{multline}
\Delta A_\text{FB}^\text{SM}(q^2) = A_\text{FB}^\text{SM}(q^2) - A_\text{FB}^\text{SM,\, no had.}(q^2)  \\
\simeq 0.86~\mathrm{Re}\,a_- + 0.10~\mathrm{Re}\,a_0 + \Big( 0.031~\mathrm{Re}\,b_- - 0.032~\mathrm{Re}\,a_0 \\ + 0.0036~\mathrm{Re}\,b_0 \Big) \frac{q^2}{\text{GeV}^2} - 0.0011~\mathrm{Re}\,b_0 \frac{q^4}{\text{GeV}^4}~,
\end{multline}
\begin{multline}
\Delta S_5^\text{SM}(q^2) = S_5^\text{SM}(q^2) - S_5^\text{SM,\, no had.}(q^2)  \\
\simeq -\bigg( 1.46 \Big( \mathrm{Re}\,a_- + \mathrm{Re}\,a_+ \Big) + 0.17~\mathrm{Re}\,a_0 \bigg) \frac{\text{GeV}}{\sqrt{q^2}} \\
-\bigg( 0.052 \Big( \mathrm{Re}\,b_- + \mathrm{Re}\,b_+ \Big)  - 0.049~\mathrm{Re}\,a_0 + 0.0061~\mathrm{Re}\,b_0 \bigg) \frac{\sqrt{q^2}}{\text{GeV}} \\ + 0.0017~\mathrm{Re}\,b_0 \frac{q^2 \sqrt{q^2}}{\text{GeV}^3} ~.
\end{multline}
Also here we stress that the expressions for $A_\text{FB}$ and $S_5$ are not expected to be very accurate. Nevertheless, they allow us to draw conclusions about the expected impact of the hadronic contributions. If the real parts of the hadronic parameters are generically as large as the imaginary parts $\mathrm{Re}\,a_0, \mathrm{Re}\,b_- \sim \mathrm{Im}\,a_0, \mathrm{Im}\,b_- \sim \mathcal O(1)$, then $A_\text{FB}$ and $S_5$ are affected at the $\mathcal O(0.1)$ level as indicated by the experimental data (c.f. figure~\ref{fig:intro}). In fact, $\mathrm{Re}\,a_0 \simeq \mathrm{Re}\,b_- \simeq -1$ mimics a new physics contribution to $\Delta C_9 \simeq -1$, which is known to be strongly preferred by the $B \to K^* \mu^+ \mu^-$ angular anomalies.

Concerning the remaining hadronic parameters that enter $A_\text{FB}$ and $S_5$: while the parameter $\mathrm{Re}\,a_-$ enters with a large prefactor, we note that it also affects the decay $B \to K^* \gamma$ and the decay $B \to K^* e^+ e^-$ at very low $q^2$. Even a very small value of $\mathrm{Re}\,a_- \sim 0.05$ corresponds to a $|m_B \mathrm{Re}\,a_- / ( 2 m_b C_7^\text{SM})| \sim 10\%$ shift in the $B \to K^* \gamma$ amplitude. Given experimental constraints~\cite{BaBar:2009byi, Belle:2017hum, LHCb:2020dof, Belle:2024mml} we expect $\mathrm{Re}\,a_-$ to play only a minor role in $A_\text{FB}$ and $S_5$. Furthermore, as already mentioned above, we expect $a_+$ and $b_+$ to be small~\cite{Beneke:2001at, Jager:2012uw} based on heavy quark expansion arguments.

\subsection{Probes of Right-Handed Currents: \texorpdfstring{$S_3$}{S3} and \texorpdfstring{$S_9$}{S9}} \label{sec:RH}

The smallness of $a_+$ and $b_+$ can in principle be checked with data on the observables $S_3$ and $S_9$.
With our approximations, we find the following expressions in the SM
\begin{multline}
S_3^\text{SM}(q^2) = \frac{16 m_{K^*}^2}{m_B^2} \frac{\xi^2_\perp(0)}{\xi^2_\parallel(0)} \bigg[\frac{m_B^2}{q^2} \left(\mathrm{Im}\,a_- +\mathrm{Im}\,b_- \frac{q^2}{m_B^2} \right)  \left(\mathrm{Im}\,a_+ + \mathrm{Im}\,b_+ \frac{q^2}{m_B^2} \right) \\
+\left( C_9^\text{SM} + \mathrm{Re}\,b_- + \frac{m_B^2}{q^2} \left( \frac{2m_b}{m_B} C_7^\text{SM} + \mathrm{Re}\,a_- \right) \right) \left(\mathrm{Re}\,a_+ + \mathrm{Re}\,b_+ \frac{q^2}{m_B^2} \right) \bigg]\frac{1}{D_\text{SM}}~,
\end{multline}
\begin{multline}
S^\text{SM}_9(q^2) = -\frac{16 m_{K^*}^2}{m_B^2} \frac{\xi^2_\perp(0)}{\xi^2_\parallel(0)} \bigg[ \frac{m_B^2}{q^2} \left(\mathrm{Im}\,a_- + \mathrm{Im}\,b_- \frac{q^2}{m_B^2} \right)  \left(\mathrm{Re}\,a_+ + \mathrm{Re}\,b_+ \frac{q^2}{m_B^2} \right) \\ - \left( C_9^\text{SM} + \mathrm{Re}\,b_- + \frac{m_B^2}{q^2} \left( \frac{2m_b}{m_B} C_7^\text{SM} + \mathrm{Re}\,a_- \right) \right) \left(\mathrm{Im}\,a_+ + \mathrm{Im}\,b_+ \frac{q^2}{m_B^2} \right) \bigg] \frac{1}{D_\text{SM}}~,
\end{multline}
with $D_\text{SM}$ given in~\eqref{eq:DSM}.
$S_3$ and $S_9$ vanish in the absence of right handed currents. In our setup, their SM predictions are therefore necessarily proportional to the real and imaginary parts of the hadronic parameters $a_+$ and $b_+$.

This is seen even more cleanly at linear order in the hadronic parameters, in which the expressions for $S_3$ and $S_9$ become approximately
\begin{eqnarray}
S_3(q^2) &\simeq& \mathrm{Re}\,a_+\left(0.59 - 1.88 ~ \frac{\text{GeV}^2}{q^2} \right) -\mathrm{Re}\,b_+\left(0.067 - 0.021 ~ \frac{q^2}{\text{GeV}^2}\right) ~, \\
S_9(q^2) &\simeq& \mathrm{Im}\,a_+\left(0.59 - 1.88 ~ \frac{\text{GeV}^2}{q^2} \right) -\mathrm{Im}\,b_+\left(0.067 - 0.021 ~ \frac{q^2}{\text{GeV}^2}\right) ~.
\end{eqnarray}
Measurements of both $S_3$ and $S_9$ are compatible with zero and have reached a precision of few $\times 10^{-2}$~\cite{LHCb:2025mqb}. This implies that the real and imaginary parts of $a_+$ and $b_+$ are limited at the level of few $\times 10^{-2}$ and few $\times 10^{-1}$, respectively.

\subsection{The direct CP Asymmetry \texorpdfstring{$A_\text{CP}$}{ACP}} \label{sec:ACP}

Next, we turn our attention to other observables that are affected by the imaginary parts of the non-local hadronic contributions. Prominent among them is the direct CP asymmetry $A_\text{CP}$, which can only be non-zero in the presence of both a strong phase and a weak phase. In our approximations, the strong phase is provided by the hadronic parameters, while the weak phase can arise from complex new physics contributions to the Wilson coefficients. We find the following expression
\begin{equation}
A_\text{CP} \simeq 2 \left( \mathrm{Im}\,a_0 + \mathrm{Im}\,b_0 \frac{q^2}{m_B^2} \right) \left( \mathrm{Im}(C_9 - C_9^\prime) + \frac{2 m_b}{m_B} \mathrm{Im}(C_7 - C_7^\prime) \right) \frac{1}{D} ~,
\end{equation}
with the denominator factor $D$ given by
\begin{multline} \label{eq:D}
 D = \big|C_{10} - C_{10}^\prime \big|^2 + \left( \frac{2m_b}{m_B} \mathrm{Re}(C_7 - C_7^\prime) + \mathrm{Re}(C_9-C_9^\prime) +\mathrm{Re}\,a_0 + \mathrm{Re}\,b_0 \frac{q^2}{m_B^2} \right)^2 \\
+ \left( \frac{2m_b}{m_B} \mathrm{Im}(C_7 - C_7^\prime) + \mathrm{Im}(C_9-C_9^\prime) \right)^2 + \left( \mathrm{Im}\,a_0 + \mathrm{Im}\,b_0 \frac{q^2}{m_B^2} \right)^2~.
\end{multline}
which generalizes the expression from equation~\eqref{eq:DSM}. Note that in the expression for $D$, the real parts of the hadronic parameters $a_0$ and $b_0$ and the Wilson coefficients $C_7^{(\prime)}$ and $C_9^{(\prime)}$ interfere with each other, but the imaginary parts do not.

The direct CP asymmetry is mainly sensitive to the imaginary parts of the hadronic parameters, $a_0$ and $b_0$. Our model describes a simple $q^2$ dependence of the direct CP asymmetry and does not capture a more pronounced $q^2$ dependence of the strong phase that can be expected closer to the $J/\psi$ resonance~\cite{Blake:2017fyh, Becirevic:2020ssj, Kamenik:2024clz}.

At linear order in the hadronic parameters and linear in the new physics contributions, our expression for $A_\text{CP}$ becomes approximately
\begin{equation}
A_\text{CP} \simeq \left( \mathrm{Im}\,a_0 + 0.036~\mathrm{Im}\,b_0 \frac{q^2}{\text{GeV}^2}\right) \Big( 0.065 ~ \mathrm{Im}\big( C_9 - C_9^\prime \big) + 0.103~\mathrm{Im}\big( C_7 - C_7^\prime \big)  \Big) ~.
\end{equation}
Measurements of the direct CP asymmetry are compatible with zero and have reached uncertainties of the order of $\pm 0.05$~\cite{LHCb:2025mqb}.
With an order one imaginary part of the hadronic parameter $a_0$, the direct CP asymmetry is sensitive to order one imaginary parts of the Wilson coefficients $C_9 - C_9^\prime$ and $C_7 - C_7^\prime$. Importantly, the sensitivity depends crucially on knowing the value of $\text{Im}\,a_0$. As argued in Section~\ref{sec:S7S8}, high precision measurements of $S_7$ and $S_8$ allow us in principle to determine $\text{Im}\, a_0$ from data, thus elevating $A_\text{CP}$ to a clean probe of CP violating new physics.

Qualitatively similar conclusions hold for other observables that are sensitive to CP violation, in particular the angular CP asymmetries $A_3$, $A_4$, $A_5$, and $A_\text{FB}^\text{CP}$. In the presence of non-zero strong phases, they become sensitive probes of the imaginary parts of $C_7^{(\prime)}$, $C_{9}^{(\prime)}$, and $C_{10}^{(\prime)}$.

\section{Numerical Results and Implications} \label{sec:numerics}

With the analytical discussion in Section~\ref{sec:analytical} suggesting sizable imaginary parts of hadronic parameters, we proceed to numerical fits of the current $B \to K^* \mu^+ \mu^-$ data.  In Section~\ref{sec:fits_CP_averaged}, we perform a fit of hadronic parameters to CP-averaged $B \to K^* \mu^+ \mu^-$ observables (including in particular $S_7$ and $S_8$) and quantify the preference for imaginary hadronic contributions. We also explore whether the associated real parts of these hadronic contributions can simultaneously resolve the $S_5$, $A_{\text{FB}}$ and branching ratio discrepancies. In Section~\ref{sec:fits_CPV}, we investigate to which extent a fit including real and imaginary parts of hadronic parameters retains sensitivity to new physics, focusing in particular on the impact of the measurements of CP asymmetries.

Our numerical analysis is based on theory predictions for the observables which we calculate with a custom version of \texttt{flavio}~\cite{Straub:2018kue}. The full dependence on the muon mass is included and the local $B \to K^*$ form factors from~\cite{Bharucha:2015bzk} including their correlated uncertainties are used (see also~\cite{Gubernari:2023puw}). The most relevant CKM input for the prediction of the $B \to K^* \mu^+ \mu^-$ branching ratio is $V_{cb}$, and we use the default \texttt{flavio} choice $|V_{cb}| = (41.97 \pm 0.48) \times 10^{-3}$~\cite{Finauri:2023kte} which is close to the current inclusive average value quoted by the PDG~\cite{ParticleDataGroup:2024cfk}.\footnote{Using instead the exclusive value $|V_{cb}| = (39.5 \pm 0.5) \times 10^{-3}$ or the PDG average $|V_{cb}| = (40.7 \pm 1.3) \times 10^{-3}$~\cite{ParticleDataGroup:2024cfk} would ease some of the tension in the $B \to K^* \mu^+\mu^-$ branching ratio~\cite{Altmannshofer:2021uub}. See~\cite{Buras:2021nns, Buras:2022qip} for an alternative, $V_{cb}$-independent approach.}

Several non-local hadronic effects are explicitly implemented in \texttt{flavio}. Leading order factorizable contributions from four-quark operators can be absorbed into effective $q^2$ dependent Wilson coefficients $C_7 \to C_7^\text{eff}$, $C_9 \to C_9^\text{eff}(q^2)$. Next-to-leading order corrections from four-quark operators and the chromo-magnetic operator can be incorporated as additional shifts to $C_7^\text{eff}$ and $C_9^\text{eff}$~\cite{Asatryan:2001zw, Beneke:2001at, Seidel:2004jh, Beneke:2004dp}. Also implemented are leading order weak annihilation contributions~\cite{Beneke:2001at}, including the power corrections from~\cite{Beneke:2004dp}, and next-to-leading order non-factorizable spectator scattering effects from four-quark operators and the chromo-magnetic operator~\cite{Beneke:2001at}.
Additional ``sub-leading'' non-local effects are parameterized exactly as in equations~\eqref{eq:h0},~\eqref{eq:hminus}, and~\eqref{eq:hplus}.

In \texttt{flavio}, the values of the hadronic parameters are Gaussian distributed and the distributions are propagated into the central values and uncertainties of the theory predictions of observables.
The default hadronic parameter ranges in \texttt{flavio v2.7} are\footnotemark
\begin{align} \label{eq:flavio1}
 \text{Re}(a_+^\texttt{flavio}) &= 0.0 \pm 0.006 ~,& \text{Re}(b_+^\texttt{flavio}) &= 0.0 \pm 0.216 ~, \\ \label{eq:flavio2}
 \text{Im}(a_+^\texttt{flavio}) &= 0.0 \pm 0.006 ~,& \text{Im}(b_+^\texttt{flavio}) &= 0.0 \pm 0.216 ~, \\ \label{eq:flavio3}
 \text{Re}(a_-^\texttt{flavio}) &= 0.0 \pm 0.023 ~,& \text{Re}(b_-^\texttt{flavio}) &= 0.0 \pm 0.432 ~, \\ \label{eq:flavio4}
 \text{Im}(a_-^\texttt{flavio}) &= 0.0 \pm 0.023 ~,& \text{Im}(b_-^\texttt{flavio}) &= 0.0 \pm 0.432 ~, \\ \label{eq:flavio5}
 \text{Re}(a_0^\texttt{flavio}) &= 0.0 \pm 0.186 ~,& \text{Re}(b_0^\texttt{flavio}) &= 0.0 \pm 2.16 ~, \\ \label{eq:flavio6}
 \text{Im}(a_0^\texttt{flavio}) &= 0.0 \pm 0.186 ~,& \text{Im}(b_0^\texttt{flavio}) &= 0.0 \pm 2.16 ~.
\end{align}
In this work, we treat $a_\lambda$ and $b_\lambda$ as free parameters and fit both hadronic parameters and new physics Wilson coefficients to the $B \to K^* \mu^+ \mu^-$ data.

\footnotetext[3]{Previous versions of \texttt{flavio} (version \texttt{v2.6.2} and earlier) used a slightly different normalization of the hadronic parameters
$$ a_\lambda^\texttt{v2.7} = \frac{2m_b}{m_B} ~a_\lambda^\texttt{v2.6.2} \simeq 1.6~a_\lambda^\texttt{v2.6.2}~,\quad b_\lambda^\texttt{v2.7} = \frac{2m_b m_B}{\text{GeV}^2} ~ b_\lambda^\texttt{v2.6.2} \simeq 43~b_\lambda^\texttt{v2.6.2}~.$$
As of version \texttt{v2.7}, \texttt{flavio} uses the normalization as in equations~\eqref{eq:h0},~\eqref{eq:hminus}, and~\eqref{eq:hplus} and the values quoted in equations~\eqref{eq:flavio1}-\eqref{eq:flavio6}.}

To handle the high-dimensional parameter space, we use the \texttt{jelli}~\cite{Smolkovic:2026cba} package, which implements automatic differentiation based on \texttt{JAX}~\cite{jax2018github} and allows constructing differentiable likelihoods that can be efficiently sampled with gradient-based methods. To this end, we express all observables as ratios of polynomials in the Wilson coefficients and hadronic parameters and store them (including their uncertainties and correlations) in the Polynomial Observable Prediction exchange format (\texttt{POPxf})~\cite{Brivio:2025mww}, which builds upon the methodology developed in~\cite{Altmannshofer:2021qrr} and is natively supported by \texttt{jelli}.
From the \texttt{POPxf} implementation of the observables and the experimental results including full correlations as provided by LHCb~\cite{LHCb:2025mqb}, we construct differentiable likelihoods with \texttt{jelli} at the renormalization scale $\mu = 4.8$~GeV. We then use Bayesian inference to estimate the posterior probability distribution of the fit parameters. This is done by Hamiltonian Monte Carlo (HMC)~\cite{Duane:1987de} sampling of the posterior using the No-U-Turn Sampler~(NUTS)~\cite{JMLR:v15:hoffman14a} implemented in the \texttt{numpyro}~\cite{phan2019composable, bingham2019pyro} package.

\subsection{Fits of Hadronic Parameters to CP-Averaged Observables} \label{sec:fits_CP_averaged}

We start with a fit of the 12 real hadronic parameters
\begin{equation} \label{eq:hadronic_parameters}
\Big\{ \mathrm{Re}\,a_\lambda ~,~~ \mathrm{Im}\,a_\lambda ~,~~ \mathrm{Re}\,b_\lambda ~,~~ \mathrm{Im}\,b_\lambda  \Big\}~, \qquad \lambda = 0,\pm ~,
\end{equation}
to the set of CP-averaged observables in $B \to K^* \mu^+ \mu^-$, keeping all Wilson coefficients fixed to their SM values.
We use two configurations of the LHCb results~\cite{LHCb:2025mqb}:
\begin{itemize}
\item {\it Configuration (vi)} which partially includes muon mass effects, uses the $S_i$ basis for CP-averaged observables, keeps CP asymmetries fixed to zero, and provides results in narrow $q^2$ bins. \item {\it Configuration (iv)} which neglects the muon mass and uses wider $q^2$ bins, but is the only configuration that provides results for both CP-averaged observables and CP asymmetries.
\end{itemize}

We restrict our fits to the low-$q^2$ region below $6\,\text{GeV}^2$, where our parameterization in equations~\eqref{eq:h0}--\eqref{eq:hplus} is expected to provide an adequate description of non-local hadronic effects.

The total number of included observables in our analysis of configuration (vi) is 72, corresponding to 8 $q^2$~bins for each of the 9 CP-averaged observables
\begin{equation}
F_L = -S_2^c ~,~~ S_3 ~,~~ S_4 ~,~~ S_5 ~,~~ A_\text{FB} ~,~~ S_7 ~,~~ S_8 ~,~~ S_9 ~,~~ \frac{d\text{BR}}{dq^2} ~.
\end{equation}
In the analysis of configuration (iv), the number of observables is 36, corresponding to 4 $q^2$~bins for each of the 9 CP-averaged observables. The CP asymmetries are not included in this part of our analysis to allow for a more direct comparison between configuration (iv) and configuration (vi), which does not provide them. Moreover, the CP asymmetries are close to zero in the absence of new physics and therefore not expected to significantly affect the fit of the hadronic parameters.

The fit results are summarized in terms of mean and standard deviation of the 12 sampled hadronic parameters in table~\ref{tab:fit}.\footnote{The notation $\widetilde{\mathrm{Re}\,a_0}$, $\widetilde{\mathrm{Re}\,b_\pm}$ in table~\ref{tab:fit} is introduced for later convenience in Section~\ref{sec:fits_CPV}. In the absence of new physics in the Wilson coefficients we have $\widetilde{\mathrm{Re}\,a_0} = \mathrm{Re}\,a_0$ and $\widetilde{\mathrm{Re}\,b_\pm} = \mathrm{Re}\,b_\pm$.}
The first column gives the results based on configuration~(vi), while column two gives the results based on configuration (iv).
The data of both configurations is able to constrain all parameters without leaving any flat directions.
The correlation matrices obtained from the sample covariance matrices are collected in tables~\ref{tab:correlation_vi_Samples} and~\ref{tab:correlation_iv_Samples} in appendix~\ref{app:correlations_had}. The correlations are very similar in both configurations.
We have checked that the maximum likelihood estimate and the inverse Hessian at the likelihood maximum give consistent central values, uncertainties, and correlations.

\renewcommand{\arraystretch}{1.0}
\begin{table}[tbh]
    \centering
    \setlength{\tabcolsep}{1pt}
    \rowcolors{3}{gray!10}{white}
       \begin{tabular}{c|c|c|c|c|c}
        \toprule
        \rowcolor{white}
        & Config (vi)
        & Config (iv)
        & \multicolumn{3}{c}{Config (iv) (CP av. + asym.)} \\
        \rowcolor{white}
        \makebox[1.5cm]{}
        & \makebox[2.9cm]{(CP av.)}
        & \makebox[2.9cm]{(CP av.)}
        & \makebox[2.9cm]{$\mathrm{Re}\,C_9^{(\prime)}$ free}
        & \makebox[2.9cm]{$\mathrm{Re}\,C_9^{(\prime)}=0$}
        & \makebox[2.9cm]{\texttt{flavio} default} \\
        \midrule
        ~$\widetilde{\mathrm{Re}\,a_0}$ & $ -0.77 \pm 0.52 $ & $ -1.04 \pm 0.52 $ & $ -0.95 \pm 0.52 $ & $ -0.97 \pm 0.52 $ & -- \\
        ~$\mathrm{Re}\,b_0$ & $ -2.45 \pm 3.59 $ & $ 1.18 \pm 3.66 $ & $ 1.50 \pm 2.98 $ & $ 1.65 \pm 3.14 $ & -- \\
        ~$\mathrm{Re}\,a_-$ & $ -0.03 \pm 0.04 $ & $ -0.06 \pm 0.04 $ & $ -0.02 \pm 0.05 $ & $ -0.02 \pm 0.05 $ & -- \\
        ~$\widetilde{\mathrm{Re}\,b_-}$ & $ -1.04 \pm 0.37 $ & $ -0.77 \pm 0.35 $ & $ -0.93 \pm 0.40 $ & $ -0.94 \pm 0.40 $ & -- \\
        ~$\mathrm{Re}\,a_+$ & $ 0.01 \pm 0.03 $ & $ 0.02 \pm 0.03 $ & $ 0.02 \pm 0.03 $ & $ 0.02 \pm 0.03 $ & -- \\
        ~$\widetilde{\mathrm{Re}\,b_+}$ & $ 0.09 \pm 0.70 $ & $ -0.15 \pm 0.61 $ & $ -0.20 \pm 0.58 $ & $ -0.20 \pm 0.59 $ & -- \\
        \midrule
        ~$\mathrm{Im}\,a_0$ & $ -0.07 \pm 0.77 $ & $ -0.90 \pm 0.78 $ & $ -0.62 \pm 0.68 $ & $ -0.67 \pm 0.73 $ & -- \\
        ~$\mathrm{Im}\,b_0$ & $ -0.38 \pm 5.62 $ & $ 4.85 \pm 6.70 $ & $ 0.87 \pm 4.94 $ & $ 1.34 \pm 5.69 $ & -- \\
        ~$\mathrm{Im}\,a_-$ & $ 0.00 \pm 0.06 $ & $ -0.01 \pm 0.07 $ & $ 0.02 \pm 0.06 $ & $ 0.02 \pm 0.06 $ & -- \\
        ~$\mathrm{Im}\,b_-$ & $ -0.58 \pm 1.00 $ & $ -0.56 \pm 0.97 $ & $ -1.03 \pm 0.80 $ & $ -1.01 \pm 0.83 $ & -- \\
        ~$\mathrm{Im}\,a_+$ & $ 0.03 \pm 0.04 $ & $ 0.04 \pm 0.04 $ & $ 0.03 \pm 0.03 $ & $ 0.03 \pm 0.03 $ & -- \\
        ~$\mathrm{Im}\,b_+$ & $ 0.42 \pm 0.76 $ & $ 0.27 \pm 0.72 $ & $ 0.58 \pm 0.62 $ & $ 0.56 \pm 0.63 $ & -- \\
        \midrule
        ~$\widetilde{\mathrm{Re}\,C_9}$ & -- & -- & $ 0.05 \pm 17.41 $ & -- & $ -0.88 \pm 0.29 $ \\
        ~$\widetilde{\mathrm{Re}\,C_9^\prime}$ & -- & -- & $ -0.64 \pm 17.21 $ & -- & $ 0.32 \pm 0.59 $ \\
        ~$\mathrm{Re}\,C_{10}$ & -- & -- & $ 0.66 \pm 0.31 $ & $ 0.65 \pm 0.32 $ & $ 0.71 \pm 0.29 $ \\
        ~$\mathrm{Re}\,C_{10}^\prime$ & -- & -- & $ 0.10 \pm 0.25 $ & $ 0.10 \pm 0.25 $ & $ -0.24 \pm 0.35 $ \\
        \midrule
        ~$\mathrm{Im}\,C_9$ & -- & -- & $ 0.39 \pm 0.46 $ & $ 0.41 \pm 0.45 $ & $ 0.12 \pm 0.43 $ \\
        ~$\mathrm{Im}\,C_9^\prime$ & -- & -- & $ 0.48 \pm 0.44 $ & $ 0.48 \pm 0.42 $ & $ 0.66 \pm 0.49 $ \\
        ~$\mathrm{Im}\,C_{10}$ & -- & -- & $ -0.31 \pm 0.28 $ & $ -0.30 \pm 0.29 $ & $ 0.18 \pm 0.24 $ \\
        ~$\mathrm{Im}\,C_{10}^\prime$ & -- & -- & $ -0.35 \pm 0.24 $ & $ -0.34 \pm 0.24 $ & $ -0.28 \pm 0.28 $ \\
        \bottomrule
    \end{tabular}%
    \caption{
    Mean values and standard deviations of sampled hadronic parameters and Wilson coefficients.
    The first two columns correspond to fits of hadronic parameters to CP-averaged observables, using LHCb configurations (vi) and (iv), respectively (see Section~\ref{sec:fits_CP_averaged}).
    The last three columns show results from fits including both CP-averaged observables and CP asymmetries using configuration~(iv). In columns three and four, hadronic parameters and Wilson coeffcients are fit simultaneously, while in column five hadronic parameters are set to \texttt{flavio} default values (see Section~\ref{sec:fits_CPV}). Dashes indicate parameters not varied in the corresponding fit.}
    \label{tab:fit}
\end{table}
\renewcommand{\arraystretch}{1.0}

Several qualitative features emerge:
\begin{itemize}
\item Overall, the results of configurations (vi) and (iv) agree well. The largest difference is in the coefficients $\mathrm{Im}\,a_0$ and $\mathrm{Im}\,b_0$ for which the central values shift by approximately $1\sigma$. We note that these two hadronic parameters are highly correlated at around $-70\%$ (see tables~\ref{tab:correlation_vi_Samples} and~\ref{tab:correlation_iv_Samples}). As both $\mathrm{Im}\,a_0$ and $\mathrm{Im}\,b_0$ affect the same amplitude but come with a different $q^2$ dependence, it appears reasonable that different $q^2$ binnings can give somewhat different best fit values for these parameters.
\item More generally, we find strong anti-correlations between the parameters $a_0$ and $b_0$, $a_-$ and $b_-$, as well as $a_+$ and $b_+$, both of their real parts and their imaginary parts. In addition, there is a non-trivial pattern of correlations across all real parts and all imaginary parts of the hadronic parameters. However, the real parts are weakly correlated with the imaginary parts.
\item The parameters $a_+$ and $b_+$ are found to be somewhat smaller compared to $a_-$ and $b_-$, in qualitative agreement with the expectation that the hadronic effects in the $\lambda=+$ helicity amplitude are suppressed in the $1/m_b$ expansion. The constraints on $a_+$ and $b_+$ are slightly weaker than naively anticipated in Section~\ref{sec:RH}.
\item As expected, there is strong preference for $\mathrm{Re}\,b_- \simeq \mathrm{Re}\,a_0 \simeq -1$, which is mainly driven by the anomalies in $S_5$, $A_\text{FB}$, and the branching ratio.
\item No single imaginary part deviates from zero by more than about $1\sigma$. Taken individually, the data therefore do not provide significant evidence for a non-vanishing strong phase in any specific helicity amplitude.
\item Nevertheless, the combination of all imaginary parts yields a substantially improved description of the data. In particular, the fitted hadronic parameters provide an excellent simultaneous description of the observables $S_7$ and $S_8$.
\end{itemize}

\begin{figure}[tb]
\centering
\includegraphics[width=0.48\textwidth]{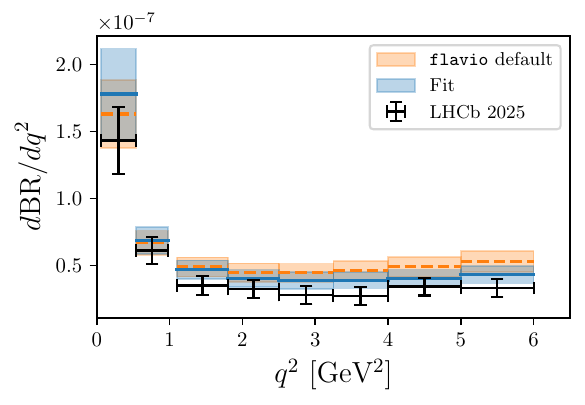} \quad
\includegraphics[width=0.48\textwidth]{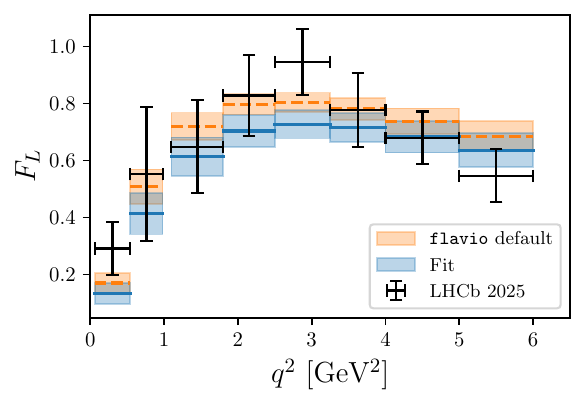} \\[8pt]
\includegraphics[width=0.48\textwidth]{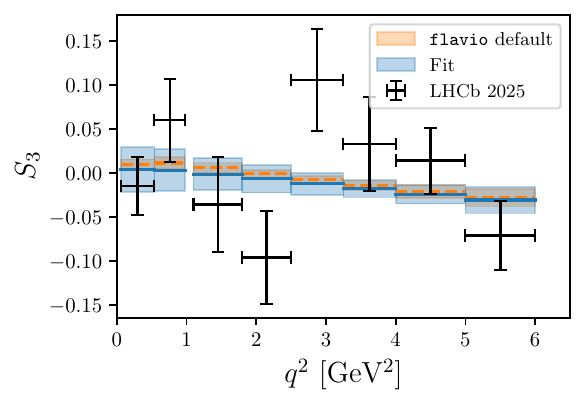} \quad
\includegraphics[width=0.48\textwidth]{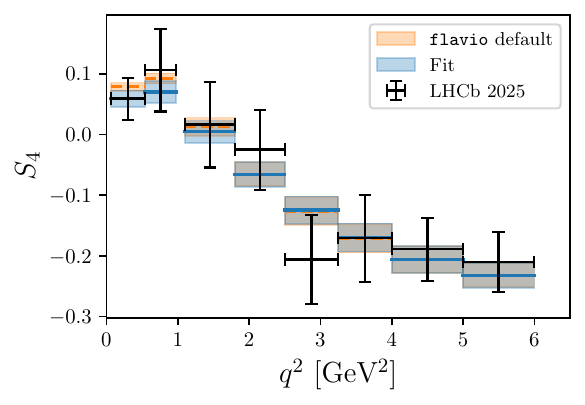}
\caption{Predictions for the $B \to K^* \mu^+ \mu^-$ branching ratio (top left) and the CP-averaged angular observables $F_L$ (top right), $S_3$ (bottom left), and $S_4$ (bottom right) in the low $q^2$ region below the $J/\psi$ resonance. Our fit predictions (blue) are compared to the default \texttt{flavio} predictions (orange) and the LHCb data (black).}
\label{fig:Si_fit_1}
\end{figure}
\begin{figure}[tb]
\centering
\includegraphics[width=0.48\textwidth]{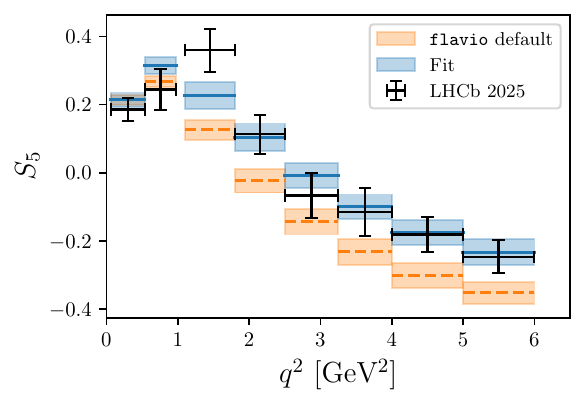} \quad
\includegraphics[width=0.48\textwidth]{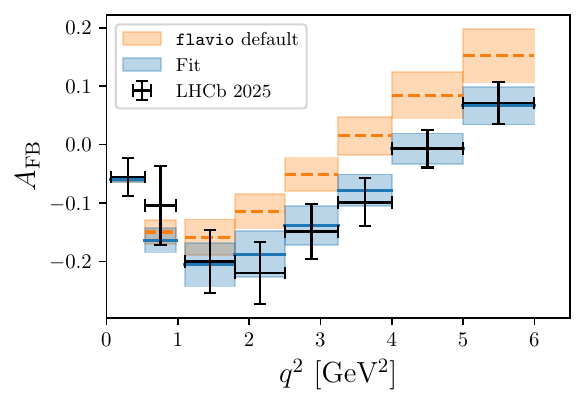} \\[8pt]
\includegraphics[width=0.48\textwidth]{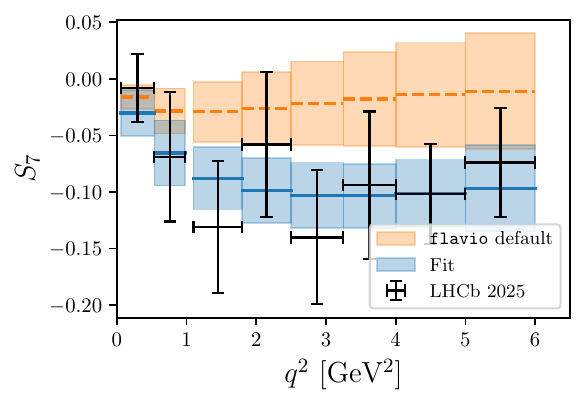} \quad
\includegraphics[width=0.48\textwidth]{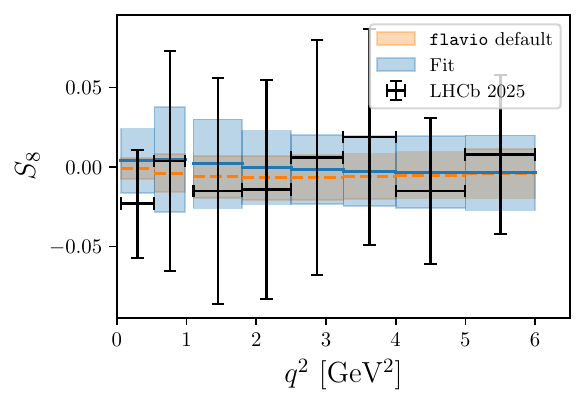} \\[8pt]
\includegraphics[width=0.48\textwidth]{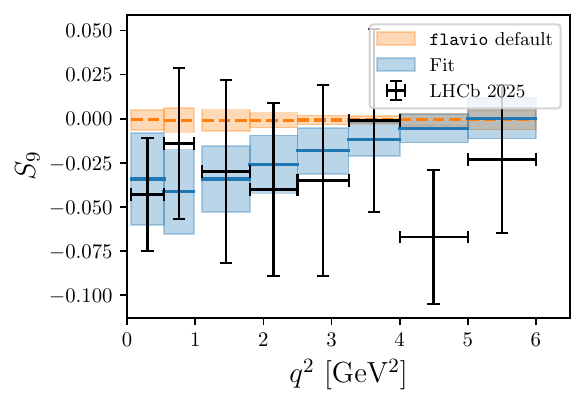}
\caption{Predictions for the CP-averaged angular observables $S_5$ (top left), $A_\text{FB}$ (top right), $S_7$ (center left), $S_8$ (center right), and $S_9$ (bottom)  in the low $q^2$ region below the $J/\psi$ resonance. Our fit predictions (blue) are compared to the default \texttt{flavio} predictions (orange) and the LHCb data (black).}
\label{fig:Si_fit_2}
\end{figure}

To illustrate the last point, we show in figures~\ref{fig:Si_fit_1} and~\ref{fig:Si_fit_2} the predictions for the CP-averaged angular observables $F_L$, $S_i$, and $A_\text{FB}$ as well as the branching ratio $d\text{BR}/dq^2$ as functions of the di-muon invariant mass squared $q^2$. We use the narrow bins of configuration (vi) and compare the fit predictions obtained from posterior samples (blue) to the default predictions of \texttt{flavio} (orange) and the LHCb data (black). The fit captures the upward shift in $S_5$, the downward shift in $A_\text{FB}$, and at least partly the overall reduction in the branching ratio. It simultaneously accommodates the deviation observed in $S_7$ while keeping $S_8$ close to zero. A small, hardly significant downward shift of $S_9$ at low $q^2$ is also fully described and is responsible for the slight preference for small non-zero imaginary parts of $a_+$ and $b_+$.

An important observation is that at the best-fit point the real and imaginary parts of several hadronic parameters are of comparable size.
This feature is not enforced by the parameterization but emerges from the fit.
At least qualitatively, this is consistent with the analytical considerations in Section~\ref{sec:analytical}, where we argued that imaginary parts large enough to explain $S_7$ naturally suggest real parts of similar size that impact $S_5$, $A_\text{FB}$, and the branching ratio at the observed level.

Interestingly, the fits are compatible with $\mathrm{Re}\,a_0 = \mathrm{Re}\,b_-$ and $\mathrm{Im}\,a_0 = \mathrm{Im}\,b_-$, i.e. an approximately $q^2$ independent complex hadronic contribution to the $\lambda = 0$ and $\lambda = -$ helicity amplitudes of equal size.

Overall, the fit demonstrates that a pattern of complex hadronic contributions can account for the full set of CP-averaged observables, in particular $S_7$ and $S_8$. So far, no single imaginary part of a hadronic parameters shows strong evidence for a non-zero value, but the fit ``dilutes'' the required imaginary contributions across many of them. If the downward shift of $S_7$ observed in the data is not a statistical fluctuation, we expect that future experimental updates might be able to establish more clearly evidence for individual imaginary parts.

\subsection{Probing CP Violating New Physics with \texorpdfstring{$B \to K^* \mu^+ \mu^-$}{B to Kstar mu mu}} \label{sec:fits_CPV}

In this section, we investigate the extent to which the data retain sensitivity to new physics once complex hadronic effects are allowed.
Because the Wilson coefficients of the dipole operators $C_7^{(\prime)}$ are best probed by $b \to s \gamma$ decays or by $b \to s e^+ e^-$ at very low $q^2$, we do not include them in our analysis. We focus instead on the following set of Wilson coefficients evaluated at the renormalization scale $\mu = 4.8$~GeV:
\begin{equation} \label{eq:WCs}
\mathrm{Re}\,C_9^{(\prime)} ~,~~
\mathrm{Im}\,C_9^{(\prime)} ~,~~
\mathrm{Re}\,C_{10}^{(\prime)} ~,~~
\mathrm{Im}\,C_{10}^{(\prime)} ~.
\end{equation}
For the fits in this section we use the LHCb results from configuration (iv), the only one in which CP asymmetries are floated. In addition to the CP-averaged observables discussed in Section~\ref{sec:fits_CP_averaged}, the fit now also includes the direct CP asymmetry $A_\text{CP}$, and the angular CP asymmetries $A_3$, $A_4$, $A_5$, $A_\text{FB}^\text{CP}$, $A_7$, $A_8$, and $A_9$. Other recent work to constrain complex Wilson coefficients using $B \to K^* \mu^+ \mu^-$ data can be found in~\cite{Biswas:2020uaq, Carvunis:2021jga, Altmannshofer:2021qrr, SinghChundawat:2022zdf, Fleischer:2022klb, Guadagnoli:2023ddc, Fleischer:2025ucq}.

\bigskip\noindent
\underline{\bf Unconstrained fit:} We start with a simultaneous fit of all hadronic parameters from equation~\eqref{eq:hadronic_parameters} and the Wilson coefficients from equations~\eqref{eq:WCs} to 68 observables from configuration~(iv), including both CP-averaged observables and CP asymmetries in the 4 wide $q^2$ bins below $6\,\text{GeV}^2$. For all fit parameters we use Gaussian priors centered at zero with standard deviation of $\pm 10$.

The real parts of the coefficients $C_9^{(\prime)}$ are largely degenerate with the parameters $\mathrm{Re}\,a_0$ and $\mathrm{Re}\,b_{\pm}$. In fact, the five parameters $\mathrm{Re}\,a_0$, $\mathrm{Re}\,b_-$, $\mathrm{Re}\,b_+$, $\mathrm{Re}\,C_9$, $\mathrm{Re}\,C_9'$ enter the approximate expressions of the helicity amplitudes in Eqs.~\eqref{eq:HVplus}, \eqref{eq:HVminus} and~\eqref{eq:HV0}, only through the following three independent combinations:
\begin{align}
    v_0 &= \mathrm{Re}\,a_0 + \mathrm{Re}\,C_9 - \mathrm{Re}\,C_9'\,, & &(H_V^0) \label{eq:v0}\\
    v_- &= \mathrm{Re}\,b_- + \mathrm{Re}\,C_9\,, & &(H_V^-) \label{eq:vm}\\
    v_+ &= \mathrm{Re}\,b_+ + \mathrm{Re}\,C_9'\,, & &(H_V^+) \label{eq:vp}
\end{align}
which span a three-dimensional subspace of the five-dimensional space spanned by $\mathrm{Re}\,a_0$, $\mathrm{Re}\,b_-$, $\mathrm{Re}\,b_+$, $\mathrm{Re}\,C_9$, $\mathrm{Re}\,C_9'$.
Within this five-dimensional space, only the three directions $v_0$, $v_-$, and $v_+$ are constrained through the (approximate) helicity amplitudes, while the two orthogonal directions
\begin{align}
    u &= \mathrm{Re}\,C_9 - \mathrm{Re}\,a_0 - \mathrm{Re}\,b_-\,,\label{eq:u}\\
    u' &= \mathrm{Re}\,C_9' + \mathrm{Re}\,a_0 - \mathrm{Re}\,b_+\,,\label{eq:uprime}
\end{align}
are unconstrained.
In the absence of NP contributions to the Wilson coefficients, i.e.\ $\mathrm{Re}\,C_9=\mathrm{Re}\,C_9'=0$, the directions $u$ and $u'$ become linear dependent on the constrained directions $v_0$, $v_-$, and $v_+$, which themselves reduce to the real parts of the hadronic parameters $a_0$, $b_-$, and $b_+$, respectively. In this case, all hadronic parameters are separately constrained. In the presence of NP, however, this is not the case anymore, and only the linear combinations in Eqs.~\eqref{eq:v0}, \eqref{eq:vm} and~\eqref{eq:vp} are constrained.

The relations in Eqs.~\eqref{eq:v0}, \eqref{eq:vm} \eqref{eq:vp}, \eqref{eq:u}, and \eqref{eq:uprime} can be summarized by
\begin{equation} \label{eq:repara_expectation}
 \begin{pmatrix}
  v_0\\
  v_-\\
  v_+\\
  u\\
  u'\\
 \end{pmatrix}
 =
 M
 \begin{pmatrix}
  \mathrm{Re}\,a_0\\
  \mathrm{Re}\,b_-\\
  \mathrm{Re}\,b_+\\
  \mathrm{Re}\,C_9\\
  \mathrm{Re}\,C_9'\\
 \end{pmatrix}
 \quad\text{with}\quad
 M =
 \begin{pmatrix}
  1 & 0 & 0 & 1 &-1\\
  0 & 1 & 0 & 1 & 0\\
  0 & 0 & 1 & 0 & 1\\
 -1 &-1 & 0 & 1 & 0\\
  1 & 0 &-1 & 0 & 1\\
 \end{pmatrix}\,,
\end{equation}
which provides a coordinate transformation between hadronic parameter and NP Wilson coefficients on one side, and the constrained and unconstrained directions on the other side.
To present our fit results, it is extremely useful to perform such a coordinate transformation to disentangle the constrained and unconstrained directions.
To improve precision beyond the heavy quark limit, as used in the derivation of Eq.~\eqref{eq:repara_expectation}, we construct a transformation matrix directly from our fit results.
To this end, we consider the $5\times 5$ sub-block of the posterior covariance matrix corresponding to the fit parameters $\mathrm{Re}\,a_0$, $\mathrm{Re}\,b_-$, $\mathrm{Re}\,b_+$, $\mathrm{Re}\,C_9$, $\mathrm{Re}\,C_9'$.
An eigendecomposition of this sub-block yields three small and two large eigenvalues, which we can identify with the three constrained and two unconstrained directions.
From linear combinations of the three constrained eigenvectors, we construct the parameters $\widetilde{\mathrm{Re}\,a_0}$, $\widetilde{\mathrm{Re}\,b_-}$, and $\widetilde{\mathrm{Re}\,b_+}$, which we require to reduce to $\mathrm{Re}\,a_0$, $\mathrm{Re}\,b_-$, and $\mathrm{Re}\,b_+$ in the limit of vanishing NP Wilson coefficients.
Similarly, from linear combinations of the unconstrained eigenvectors we construct the parameters $\widetilde{\mathrm{Re}\,C_9}$ and $\widetilde{\mathrm{Re}\,C_9'}$, which we require to reduce to $\mathrm{Re}\,C_9$ and $\mathrm{Re}\,C'_9$ in the limit of vanishing hadronic parameters.
The resulting coordinate transformation is given by
\begin{equation} \label{eq:repara}
 \begin{pmatrix}
  \widetilde{\mathrm{Re}\,a_0}\\
  \widetilde{\mathrm{Re}\,b_-}\\
  \widetilde{\mathrm{Re}\,b_+}\\
  \widetilde{\mathrm{Re}\,C_9}\\
  \widetilde{\mathrm{Re}\,C_9'}\\
 \end{pmatrix}
 =
 \widetilde M
 \begin{pmatrix}
  \mathrm{Re}\,a_0\\
  \mathrm{Re}\,b_-\\
  \mathrm{Re}\,b_+\\
  \mathrm{Re}\,C_9\\
  \mathrm{Re}\,C_9'\\
 \end{pmatrix}
 \quad\text{with}\quad
 \widetilde M \approx
 \begin{pmatrix}
   1 & 0 & 0 & 0.93 &-0.93\\
   0 & 1 & 0 & 1.09 &-0.08\\
   0 & 0 & 1 &-0.08 & 1.09\\
   -0.93 &-1.09 & 0.08 & 1 & 0\\
    0.93 & 0.08 &-1.09 & 0 & 1\\
 \end{pmatrix}\,.
\end{equation}
The $\mathcal O(10\%)$ difference between $\widetilde M$ in Eq.~\eqref{eq:repara} and $M$ in Eq.~\eqref{eq:repara_expectation} arises because in our numerical analysis we use local form factors beyond the heavy quark limit. Therefore, the linear combinations that enter the helicity amplitudes are in fact slightly different from~\eqref{eq:v0}, \eqref{eq:vm} and~\eqref{eq:vp}, and also change slightly with $q^2$.

We show the results of the fit in the third row of table~\ref{tab:fit}. Analogously to the fits of the hadronic parameters only, we derive the central values and $1\sigma$ uncertainties of all fit parameters from the mean and variance of samples. We have checked that using the inverse Hessian of the likelihood at the best fit points gives results that are fully consistent within the uncertainties. We do not observe any sign of large non-Gaussian features in the likelihood.

Comparing to the fits from Section~\ref{sec:fits_CP_averaged}, we see that the results for the hadronic parameters are largely compatible. Some of the central values shift, but typically well within the uncertainties. The anomalies in $S_5$ and $A_\text{FB}$ point to $\widetilde{\text{Re}\,a_0} \simeq \widetilde{\text{Re}\,b_-} \simeq -1$ but cannot distinguish between hadronic effects in $\text{Re}\,a_0$ and $\text{Re}\,b_-$ on the one side or a new physics contribution to $\text{Re}\,C_9$ on the other. The parameter combinations $\widetilde{\text{Re}\,C_9}$ and $\widetilde{\text{Re}\,C_9^\prime}$ are unconstrained. In fact, the $1\sigma$ uncertainties of approximately $17 \simeq \sqrt{3} \times 10$ correspond precisely to the prior range of the linear combinations that define $\widetilde{\text{Re}\,C_9}$ and $\widetilde{\text{Re}\,C_9^\prime}$ in equation~\eqref{eq:repara}.

\begin{figure}[tb]
\centering
\includegraphics[width=\textwidth]{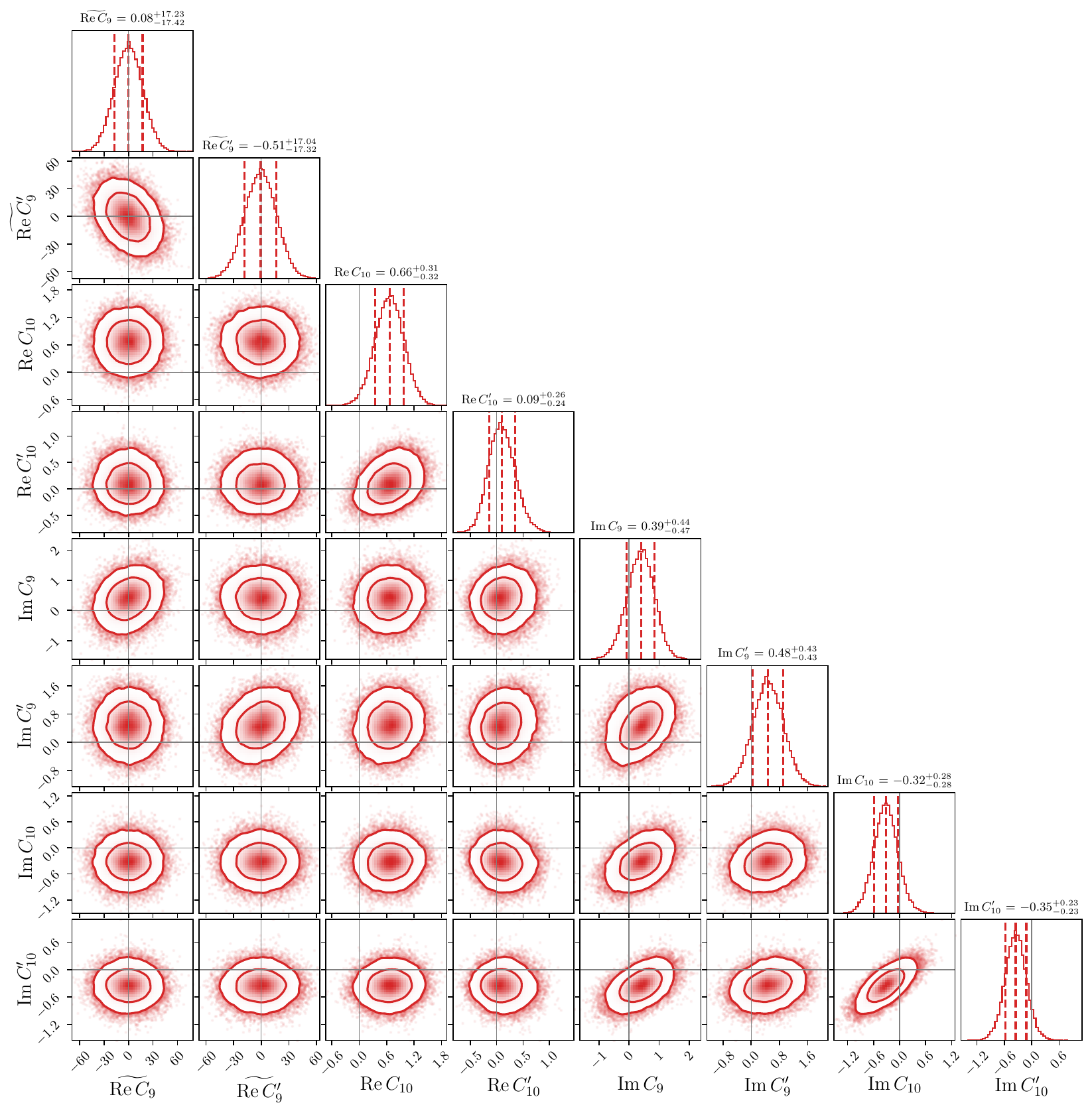}
\caption{Corner plot showing the Wilson coefficients in the combined fit of hadronic parameters and Wilson coefficients to the CP-averaged observables and CP asymmetries from LHCb configuration (iv). The diagonal panels show the one-dimensional marginalized distributions for each Wilson coefficient, while the off-diagonal panels show the corresponding two-dimensional joint distributions for all Wilson coefficient pairs with $1\sigma$ and $2\sigma$ contours.}
\label{fig:WC_contours}
\end{figure}

The fit is able to well constrain all other Wilson coefficients.
The imaginary parts of $C_9$ and $C_9^\prime$ as well as both real and imaginary parts of $C_{10}$ and $C_{10}^\prime$ have fairly small uncertainties of around $0.2$ to $0.4$. None of the imaginary parts differs significantly from zero. Interestingly, there is a $\sim 2\sigma$ preference for a small positive shift of $\text{Re}\,C_{10}$ which helps to bring the predicted $B \to K^* \mu^+ \mu^-$ branching ratio in agreement with the data.\footnote{Note that in our fit we have not included the leptonic decay $B_s \to \mu^+ \mu^-$ which is known as a clean probe of $\text{Re}\,C_{10}$. An updated global study of rare $b \to s \ell \ell$ transition is left for future work.}

To visualize the structure of the sampled new physics parameter space, we show marginal distributions of the Wilson coefficients as a ``corner plot'' in figure~\ref{fig:WC_contours}. The diagonal panels show the one-dimensional marginalized distributions for the real and imaginary parts of each Wilson coefficient, while the off-diagonal panels show the corresponding two-dimensional joint distributions for all Wilson coefficient pairs. The two-dimensional plots show $68\%$ and $95\%$ contours. Note that the values for the Wilson coefficients that are quoted in the corner plot correspond to the median and the 68\% percentile interval. They are consistent (but don't coincide exactly) with the values in table~\ref{tab:fit}, that correspond to the mean and the standard deviation.

All the one-dimensional distributions appear to very good approximation Gaussian. Also in the two dimensional distributions we do not observe any pronounced non-Gaussian features. We find a strong correlation between the imaginary parts of $C_{10}$ and $C_{10}^\prime$, which indicates that it is mainly the combination $\text{Im}\,C_{10} - \text{Im}\,C_{10}^\prime$ which is constrained by measurements of CP asymmetries.

The corresponding correlation matrix of the Wilson coefficients is reported in table~\ref{tab:correlation_np_reparam_Samples} in appendix~\ref{app:correlations_WC}. The matrix clearly shows the same features that can be seen in the corner plot discussed above.

\bigskip\noindent
\underline{\bf Comparison to alternative approaches:}
Finally, we compare the results of the unconstrained fit to two alternative approaches.
\begin{itemize}
\item[(a)] We simultaneously fit all hadronic parameters and the Wilson coefficients but switch off new physics in the real parts of $C_9$ and $C_9^\prime$ by imposing by hand $\text{Re}\,C_9 = \text{Re}\,C_9^\prime = 0$. In this approach $\widetilde{\text{Re}\,a_0}$ and $\widetilde{\text{Re}\,b_\pm}$ are simply given by the hadronic parameters $\text{Re}\,a_0$ and $\text{Re}\,b_\pm$. The coefficients $\widetilde{\text{Re}\,C_9}$ and $\widetilde{\text{Re}\,C_9^\prime}$ are not independent parameters and therefore not included in the fit. The corresponding fit results are shown in the fourth column of table~\ref{tab:fit}.
\item[(b)] We fit only the new physics Wilson coefficients and treat the hadronic parameters as nuisance parameters with their values set to the \texttt{flavio} defaults quoted in equations~\eqref{eq:flavio1}-\eqref{eq:flavio6}. In this approach, $\widetilde{\text{Re}\,C_9}$ and $\widetilde{\text{Re}\,C_9^\prime}$ should be interpreted as the new physics Wilson coefficients $\text{Re}\,C_9$ and $\text{Re}\,C_9^\prime$. The corresponding fit results are reported in the last column of table~\ref{tab:fit}.
\end{itemize}

The correlation matrices for the Wilson coefficients in the additional fits are given in tables~\ref{tab:correlation_np_zero_c9_c9p_Samples} and~\ref{tab:correlation_np_flavio_Samples} in appendix~\ref{app:correlations_WC}.

Comparing the fit in which we set $\text{Re}\,C_9$ and $\text{Re}\,C_9^\prime$ to zero to the unconstrained fit in which we allow them to float (column four vs. column three in table~\ref{tab:fit}), we see that the results for the hadronic parameters are fairly stable. The same is true for the Wilson coefficients which hardly shift between the two approaches.

As expected, the ``\texttt{flavio} default fit'' shows strong preference for $\text{Re}\,C_9 \simeq -1$. While the unconstrained fit cannot distinguish $\text{Re}\,C_9$ from $\text{Re}\,a_0 \simeq \text{Re}\,b_-$. We observe reasonably good agreement among all the central values of the other Wilson coefficients within uncertainties (column five vs. column three in table~\ref{tab:fit}). The largest shift is in $\text{Im}\,C_{10}$ with a central value that moves by approximately $1.5\sigma$.

In this context it is worth commenting on the role of the direct CP asymmetry $A_\text{CP}$ and other angular CP asymmetries like $A_3$, $A_4$, $A_5$, and $A_\text{FB}^\text{CP}$. Na{\"i}vely, as argued in Section~\ref{sec:ACP}, one expects that the presence of sizable strong phases would enhance the sensitivity of these CP asymmetries to imaginary parts of Wilson coefficients. In practice, however, the fits in which the hadronic parameters are determined by data do currently not yield significantly stronger constraints on imaginary parts of the Wilson coefficients than the fit in which the hadronic parameters are fixed to their \texttt{flavio} default values, where their central values vanish. This behavior is likely related to the fact that the present fits do not yet show clear evidence for sizable imaginary parts in individual hadronic parameters. As discussed above, the fit tends to distribute the required strong phase among several parameters, which dilutes the sensitivity to CP-violating new physics contributions.

\bigskip
As a final comment, we re-iterate that the posterior distributions resulting from our fits are all close to Gaussian. This allows for a simple and efficient reuse of our results: the combination of mean values, standard deviations, and correlation matrices defines multivariate Gaussian distributions that provide a good approximation to the full posterior, both for the hadronic parameters and the Wilson coefficients. The necessary inputs are given in table~\ref{tab:fit} (mean values and standard deviations) and tables~\ref{tab:correlation_vi_Samples} – \ref{tab:correlation_np_flavio_Samples} (correlations), enabling their use in subsequent phenomenological analyses.

\section{Summary and Outlook} \label{sec:conclusion}

The recent LHCb update of the angular analysis of $B \to K^* \mu^+\mu^-$ provides an intriguing new piece of information about the long-standing anomalies in this decay. In addition to confirming the deviations in $S_5$ and $A_\text{FB}$, the new measurements hint at a systematic downward shift in the observable $S_7$ across the low-$q^2$ region. As we have emphasized, the observables $S_7$ and $S_8$ play a unique role among the angular coefficients. Being proportional to strong phases, non-zero values of $S_7$ or $S_8$ cannot be generated by heavy new physics and instead require a sizable hadronic effect that introduces an absorptive contribution to the decay amplitudes. The observed central values of $S_7$ indicate an hadronic effect of approximately the same size required to explain $S_5$ and $A_\text{FB}$. If the emerging pattern in the data is confirmed by more precise measurements, it would provide strong evidence for unexpectedly large non-local hadronic contributions in $B \to K^* \mu^+\mu^-$.

Motivated by this observation, we have investigated a simple phenomenological description of non-local hadronic contributions and confronted it with the available LHCb data. Fits to the hadronic parameters show that the data can be described very well within this framework, see figures~\ref{fig:Si_fit_1} and~\ref{fig:Si_fit_2}. At the same time, the fits do not give clear preference for a large imaginary part in any individual hadronic parameter. Instead, the fits tends to distribute the required imaginary parts among several of them. Still, at the best-fit point, the real and imaginary parts of several hadronic parameters are of comparable size. Interestingly, the fits are compatible with an approximately $q^2$ independent complex hadronic contribution to the $\lambda = 0$ and $\lambda = -$ helicity amplitudes of equal size. We have verified that this behavior is robust with respect to different implementations of the LHCb measurements used in the fit.

We have also explored the interplay between hadronic effects and possible new physics contributions by performing combined fits of hadronic parameters and short-distance Wilson coefficients. In a fully unconstrained setup, where all hadronic parameters and Wilson coefficients are allowed to float simultaneously, the fit exhibits two flat directions in parameter space, involving the Wilson coefficients $\text{Re}\,C_9$ and $\text{Re}\,C_9^\prime$. The other Wilson coefficients $\text{Im}\,C_9$, $\text{Im}\,C_9^\prime$, $\text{Re}\,C_{10}$, $\text{Re}\,C_{10}^\prime$, $\text{Im}\,C_{10}$, and $\text{Im}\,C_{10}^\prime$ can be all constrained simultaneously.
Imposing the condition $\mathrm{Re}\,C_9=\mathrm{Re}\,C_9^\prime=0$ removes the flat directions and gives results that are fully consistent with the unconstrained fit. A fit in which the hadronic parameters are fixed to their default values in \texttt{flavio} and only the Wilson coefficients are varied, also yields results that are broadly consistent with those of the unconstrained fit. Even in the presence of sizable complex hadronic contributions and parameter degeneracies, the data robustly retains sensitivity to all Wilson coefficients other than $\mathrm{Re}\,C_9$ and $\mathrm{Re}\,C_9^\prime$.

Na{\"i}vely, one expects that in the presence of sizable strong phases, the direct CP asymmetry $A_\text{CP}$ and certain angular CP asymmetries like $A_3$, $A_4$, $A_5$, and $A_\text{FB}^\text{CP}$ show enhanced sensitivity to weak phases coming from complex Wilson coefficients.
In practice, we find that using the current data to determine the imaginary parts of the hadronic parameters does not yet improve the sensitivity compared to the fit with \texttt{flavio} default hadronic parameters.

Looking ahead, improved experimental precision will likely be crucial for clarifying the situation. If the current hint for a non-zero $S_7$ persists with additional data, fits of the hadronic parameters should begin to exhibit clearer evidence for non-zero imaginary parts, which would constitute the only possible origin of the required strong phase in the framework considered here. More precise measurements of the angular observables and CP asymmetries should also help to further improve the determination of the Wilson coefficients $\text{Im}\,C_9$, $\text{Im}\,C_9^\prime$, $\text{Re}\,C_{10}$, $\text{Re}\,C_{10}^\prime$, $\text{Im}\,C_{10}$, and $\text{Im}\,C_{10}^\prime$.

At the same time, it will be essential to better understand the origin of the large hadronic contributions suggested by the data. At present, such sizable effects are not supported by model calculations~\cite{Khodjamirian:2010vf, Gubernari:2020eft, Mahajan:2024xpo, Isidori:2024lng, Isidori:2025dkp} (see also~\cite{Ladisa:2022vmh}) and might be in tension with arguments based on unitarity and analyticity~\cite{Gubernari:2020eft, Gubernari:2022hxn, Mutke:2024tww, Gopal:2024mgb}.

A natural next step will be to extend our analysis to a more comprehensive global fit including additional rare $B$ decay observables. In particular, it will be important to incorporate information from $B \to K^*\gamma$ and $B \to K^* e^+e^-$ that depend on the same hadronic parameters. Further constraints arise from  $B \to K^* \mu^+\mu^-$ in the high-$q^2$ region and related FCNC modes including $B_s \to \mu^+\mu^-$, $B \to K \mu^+\mu^-$, and $B_s \to \phi \mu^+\mu^-$ even though they introduce dependence on additional hadronic parameters. Our fitting framework is sufficiently flexible and efficient to accommodate the large number of additional observables and hadronic parameters required for such an analysis. A global fit of this kind could already lead to improved constraints on both hadronic contributions and possible new physics effects. The results will be presented elsewhere.

\section*{Acknowledgements}

We thank Christoph Langenbruch for useful discussion concerning the results of~\cite{LHCb:2025mqb}.
The research of WA and SC is supported by the U.S. Department of Energy grant number DE-SC0010107.

\begin{appendix}
\section{Approximate New Physics Expressions for \texorpdfstring{$S_7$}{S7} and \texorpdfstring{$S_8$}{S8}} \label{app:NP}

In this appendix we present simple approximate expressions for the angular observables $S_7$ and $S_8$ in the presence of new physics in the Wilson coefficients $C_7$, $C_9$, $C_{10}$ and their primed counterparts. We use local form factors in the heavy quark limit as in equation~\eqref{eq:ff} and the simple hadronic model from equations~\eqref{eq:h0},~\eqref{eq:hminus} and~\eqref{eq:hplus}. We find
\begin{multline} \label{eq:S7_NP}
S_7^\text{SM}(q^2) \simeq -\frac{4 m_{K^*} \sqrt{q^2}}{m_B^2} \frac{\xi_\perp(0)}{\xi_\parallel(0)} \mathrm{Re}(C_{10}  - C_{10}^\prime) \bigg(\frac{m_B^2}{q^2} \Big( \mathrm{Im}\,a_-  - \mathrm{Im}\,a_+ \Big) \\ + \mathrm{Im}\,b_-  - \mathrm{Im}\,b_+ - \mathrm{Im}\,a_0 - \mathrm{Im}\,b_0 \frac{q^2}{m_B^2} \bigg) \frac{1}{D} ~,
\end{multline}
\begin{multline} \label{eq:S8_NP}
S_8^\text{SM}(q^2) \simeq \frac{2 m_{K^*} \sqrt{q^2}}{m_B^2} \frac{\xi_\perp(0)}{\xi_\parallel(0)} \Bigg[ \left(\frac{2m_b}{m_B}\mathrm{Re}(C_7 - C_7^\prime) + \mathrm{Re}(C_9 - C_9^\prime) + \mathrm{Re}\,a_0 + \mathrm{Re}\,b_0 \frac{q^2}{m_B^2}  \right) \\
\times \bigg(\frac{m_B^2}{q^2} \Big( \mathrm{Im}\,a_- + \mathrm{Im}\,a_+ \Big) + \mathrm{Im}\,b_- + \mathrm{Im}\,b_+ \bigg) \\
-  \left(\mathrm{Re}(C_9 + C_9^\prime)+ \mathrm{Re}\,b_- + \mathrm{Re}\,b_+ + \frac{ m_B^2}{q^2} \left( \frac{2m_b}{m_B} \mathrm{Re}(C_7 + C_7^\prime)+ \mathrm{Re}\,a_- + \mathrm{Re}\,a_+ \right) \right) \\ \times  \left(\mathrm{Im}\,a_0 + \mathrm{Im}\,b_0 \frac{q^2}{m_B^2} \right)  \Bigg] \frac{1}{D}~,
\end{multline}
with $D$ given in equation~\eqref{eq:D}. As in the SM case discussed in Section~\ref{sec:S7S8}, imaginary parts of the hadronic parameters are required for $S_7$ and $S_8$ to take non-zero values.

\section{Correlation Matrices of the Hadronic Fit} \label{app:correlations_had}

In this appendix, we collect the correlation matrices of the 12 hadronic parameters obtained from the fit to CP-averaged observables in Section~\ref{sec:fits_CP_averaged}. Table~\ref{tab:correlation_vi_Samples} contains the results using LHCb configuration (vi).
The corresponding correlation matrix using LHCb configuration (iv) is shown in table~\ref{tab:correlation_iv_Samples}.
Both matrices are obtain from samples of the likelihood. We have checked that using instead the inverse Hessian of the likelihood at the best fit point gives consistent results.

\begin{table}[tbh]
    \centering
    \small
    \setlength{\tabcolsep}{1.4pt}
        \begin{tabular}{l|cccccc|cccccc}
        \toprule
          & $\widetilde{\mathrm{Re}\,a_0}$ & $\mathrm{Re}\,b_0$ & $\mathrm{Re}\,a_-$ & $\widetilde{\mathrm{Re}\,b_-}$ & $\mathrm{Re}\,a_+$ & $\widetilde{\mathrm{Re}\,b_+}$ & $\mathrm{Im}\,a_0$ & $\mathrm{Im}\,b_0$ & $\mathrm{Im}\,a_-$ & $\mathrm{Im}\,b_-$ & $\mathrm{Im}\,a_+$ & $\mathrm{Im}\,b_+$ \\
        \midrule
        $\widetilde{\mathrm{Re}\,a_0}$ & \cellcolor{black!34}1.00 & \cellcolor{black!16}-0.48 & \cellcolor{black!8}0.23 & \cellcolor{black!7}-0.21 & \cellcolor{black!5}-0.16 & \cellcolor{black!9}-0.27 & \cellcolor{black!3}-0.09 & \cellcolor{black!0}-0.03 & \cellcolor{black!0}0.03 & \cellcolor{black!4}-0.14 & \cellcolor{black!1}-0.03 & \cellcolor{black!2}0.06 \\
        $\mathrm{Re}\,b_0$ & \cellcolor{black!16}-0.48 & \cellcolor{black!34}1.00 & \cellcolor{black!9}-0.27 & \cellcolor{black!7}0.20 & \cellcolor{black!11}0.34 & \cellcolor{black!15}-0.45 & \cellcolor{black!0}-0.02 & \cellcolor{black!0}-0.03 & \cellcolor{black!2}0.07 & \cellcolor{black!5}-0.15 & \cellcolor{black!1}0.05 & \cellcolor{black!1}-0.04 \\
        $\mathrm{Re}\,a_-$ & \cellcolor{black!8}0.23 & \cellcolor{black!9}-0.27 & \cellcolor{black!34}1.00 & \cellcolor{black!26}-0.76 & \cellcolor{black!3}-0.09 & \cellcolor{black!2}0.06 & \cellcolor{black!1}0.03 & \cellcolor{black!2}0.07 & \cellcolor{black!1}-0.03 & \cellcolor{black!4}0.12 & \cellcolor{black!0}-0.01 & \cellcolor{black!2}-0.06 \\
        $\widetilde{\mathrm{Re}\,b_-}$ & \cellcolor{black!7}-0.21 & \cellcolor{black!7}0.20 & \cellcolor{black!26}-0.76 & \cellcolor{black!35}1.00 & \cellcolor{black!5}0.17 & \cellcolor{black!6}-0.19 & \cellcolor{black!1}-0.03 & \cellcolor{black!2}-0.07 & \cellcolor{black!4}0.12 & \cellcolor{black!7}-0.21 & \cellcolor{black!2}0.07 & \cellcolor{black!2}-0.06 \\
        $\mathrm{Re}\,a_+$ & \cellcolor{black!5}-0.16 & \cellcolor{black!11}0.34 & \cellcolor{black!3}-0.09 & \cellcolor{black!5}0.17 & \cellcolor{black!35}1.00 & \cellcolor{black!21}-0.60 & \cellcolor{black!0}-0.00 & \cellcolor{black!2}-0.06 & \cellcolor{black!6}0.18 & \cellcolor{black!6}-0.19 & \cellcolor{black!2}0.07 & \cellcolor{black!1}-0.05 \\
        $\widetilde{\mathrm{Re}\,b_+}$ & \cellcolor{black!9}-0.27 & \cellcolor{black!15}-0.45 & \cellcolor{black!2}0.06 & \cellcolor{black!6}-0.19 & \cellcolor{black!21}-0.60 & \cellcolor{black!34}1.00 & \cellcolor{black!2}0.08 & \cellcolor{black!2}0.08 & \cellcolor{black!6}-0.18 & \cellcolor{black!12}0.36 & \cellcolor{black!2}-0.08 & \cellcolor{black!2}0.07 \\
        \midrule
        $\mathrm{Im}\,a_0$ & \cellcolor{black!3}-0.09 & \cellcolor{black!0}-0.02 & \cellcolor{black!1}0.03 & \cellcolor{black!1}-0.03 & \cellcolor{black!0}-0.00 & \cellcolor{black!2}0.08 & \cellcolor{black!35}1.00 & \cellcolor{black!24}-0.69 & \cellcolor{black!2}0.08 & \cellcolor{black!3}0.09 & \cellcolor{black!5}-0.16 & \cellcolor{black!2}-0.07 \\
        $\mathrm{Im}\,b_0$ & \cellcolor{black!0}-0.03 & \cellcolor{black!0}-0.03 & \cellcolor{black!2}0.07 & \cellcolor{black!2}-0.07 & \cellcolor{black!2}-0.06 & \cellcolor{black!2}0.08 & \cellcolor{black!24}-0.69 & \cellcolor{black!35}1.00 & \cellcolor{black!12}-0.35 & \cellcolor{black!13}0.39 & \cellcolor{black!8}0.25 & \cellcolor{black!10}-0.31 \\
        $\mathrm{Im}\,a_-$ & \cellcolor{black!0}0.03 & \cellcolor{black!2}0.07 & \cellcolor{black!1}-0.03 & \cellcolor{black!4}0.12 & \cellcolor{black!6}0.18 & \cellcolor{black!6}-0.18 & \cellcolor{black!2}0.08 & \cellcolor{black!12}-0.35 & \cellcolor{black!35}1.00 & \cellcolor{black!27}-0.78 & \cellcolor{black!12}0.34 & \cellcolor{black!4}-0.12 \\
        $\mathrm{Im}\,b_-$ & \cellcolor{black!4}-0.14 & \cellcolor{black!5}-0.15 & \cellcolor{black!4}0.12 & \cellcolor{black!7}-0.21 & \cellcolor{black!6}-0.19 & \cellcolor{black!12}0.36 & \cellcolor{black!3}0.09 & \cellcolor{black!13}0.39 & \cellcolor{black!27}-0.78 & \cellcolor{black!35}1.00 & \cellcolor{black!9}-0.27 & \cellcolor{black!6}0.18 \\
        $\mathrm{Im}\,a_+$ & \cellcolor{black!1}-0.03 & \cellcolor{black!1}0.05 & \cellcolor{black!0}-0.01 & \cellcolor{black!2}0.07 & \cellcolor{black!2}0.07 & \cellcolor{black!2}-0.08 & \cellcolor{black!5}-0.16 & \cellcolor{black!8}0.25 & \cellcolor{black!12}0.34 & \cellcolor{black!9}-0.27 & \cellcolor{black!35}1.00 & \cellcolor{black!22}-0.65 \\
        $\mathrm{Im}\,b_+$ & \cellcolor{black!2}0.06 & \cellcolor{black!1}-0.04 & \cellcolor{black!2}-0.06 & \cellcolor{black!2}-0.06 & \cellcolor{black!1}-0.05 & \cellcolor{black!2}0.07 & \cellcolor{black!2}-0.07 & \cellcolor{black!10}-0.31 & \cellcolor{black!4}-0.12 & \cellcolor{black!6}0.18 & \cellcolor{black!22}-0.65 & \cellcolor{black!35}1.00 \\
        \bottomrule
    \end{tabular}
    \caption{Correlation matrix of the 12 hadronic parameters obtained from the fit to CP-averaged observables using measurements of LHCb configuration (vi) and sampling the likelihood.}
    \label{tab:correlation_vi_Samples}
\end{table}
\begin{table}[tbh]
    \centering
    \small
    \setlength{\tabcolsep}{1.4pt}
        \begin{tabular}{l|cccccc|cccccc}
        \toprule
          & $\widetilde{\mathrm{Re}\,a_0}$ & $\mathrm{Re}\,b_0$ & $\mathrm{Re}\,a_-$ & $\widetilde{\mathrm{Re}\,b_-}$ & $\mathrm{Re}\,a_+$ & $\widetilde{\mathrm{Re}\,b_+}$ & $\mathrm{Im}\,a_0$ & $\mathrm{Im}\,b_0$ & $\mathrm{Im}\,a_-$ & $\mathrm{Im}\,b_-$ & $\mathrm{Im}\,a_+$ & $\mathrm{Im}\,b_+$ \\
        \midrule
        $\widetilde{\mathrm{Re}\,a_0}$ & \cellcolor{black!35}1.00 & \cellcolor{black!21}-0.60 & \cellcolor{black!7}0.22 & \cellcolor{black!7}-0.21 & \cellcolor{black!7}-0.22 & \cellcolor{black!4}-0.14 & \cellcolor{black!0}0.02 & \cellcolor{black!2}0.07 & \cellcolor{black!3}-0.09 & \cellcolor{black!2}0.07 & \cellcolor{black!0}-0.01 & \cellcolor{black!1}-0.06 \\
        $\mathrm{Re}\,b_0$ & \cellcolor{black!21}-0.60 & \cellcolor{black!35}1.00 & \cellcolor{black!7}-0.23 & \cellcolor{black!5}0.14 & \cellcolor{black!10}0.30 & \cellcolor{black!14}-0.41 & \cellcolor{black!4}-0.14 & \cellcolor{black!4}-0.12 & \cellcolor{black!5}0.17 & \cellcolor{black!9}-0.27 & \cellcolor{black!1}-0.03 & \cellcolor{black!7}0.22 \\
        $\mathrm{Re}\,a_-$ & \cellcolor{black!7}0.22 & \cellcolor{black!7}-0.23 & \cellcolor{black!34}1.00 & \cellcolor{black!25}-0.71 & \cellcolor{black!3}-0.10 & \cellcolor{black!2}0.07 & \cellcolor{black!3}-0.10 & \cellcolor{black!9}0.26 & \cellcolor{black!5}-0.17 & \cellcolor{black!8}0.26 & \cellcolor{black!1}0.03 & \cellcolor{black!5}-0.17 \\
        $\widetilde{\mathrm{Re}\,b_-}$ & \cellcolor{black!7}-0.21 & \cellcolor{black!5}0.14 & \cellcolor{black!25}-0.71 & \cellcolor{black!34}1.00 & \cellcolor{black!5}0.17 & \cellcolor{black!7}-0.21 & \cellcolor{black!3}0.11 & \cellcolor{black!9}-0.28 & \cellcolor{black!8}0.24 & \cellcolor{black!11}-0.34 & \cellcolor{black!0}-0.00 & \cellcolor{black!3}0.09 \\
        $\mathrm{Re}\,a_+$ & \cellcolor{black!7}-0.22 & \cellcolor{black!10}0.30 & \cellcolor{black!3}-0.10 & \cellcolor{black!5}0.17 & \cellcolor{black!35}1.00 & \cellcolor{black!21}-0.60 & \cellcolor{black!0}0.02 & \cellcolor{black!4}-0.11 & \cellcolor{black!7}0.22 & \cellcolor{black!7}-0.22 & \cellcolor{black!0}0.02 & \cellcolor{black!1}0.04 \\
        $\widetilde{\mathrm{Re}\,b_+}$ & \cellcolor{black!4}-0.14 & \cellcolor{black!14}-0.41 & \cellcolor{black!2}0.07 & \cellcolor{black!7}-0.21 & \cellcolor{black!21}-0.60 & \cellcolor{black!34}1.00 & \cellcolor{black!2}0.07 & \cellcolor{black!3}0.10 & \cellcolor{black!6}-0.17 & \cellcolor{black!9}0.27 & \cellcolor{black!0}-0.01 & \cellcolor{black!2}-0.08 \\
        \midrule
        $\mathrm{Im}\,a_0$ & \cellcolor{black!0}0.02 & \cellcolor{black!4}-0.14 & \cellcolor{black!3}-0.10 & \cellcolor{black!3}0.11 & \cellcolor{black!0}0.02 & \cellcolor{black!2}0.07 & \cellcolor{black!35}1.00 & \cellcolor{black!24}-0.70 & \cellcolor{black!6}0.17 & \cellcolor{black!2}-0.07 & \cellcolor{black!5}-0.16 & \cellcolor{black!1}-0.03 \\
        $\mathrm{Im}\,b_0$ & \cellcolor{black!2}0.07 & \cellcolor{black!4}-0.12 & \cellcolor{black!9}0.26 & \cellcolor{black!9}-0.28 & \cellcolor{black!4}-0.11 & \cellcolor{black!3}0.10 & \cellcolor{black!24}-0.70 & \cellcolor{black!35}1.00 & \cellcolor{black!15}-0.44 & \cellcolor{black!19}0.55 & \cellcolor{black!9}0.27 & \cellcolor{black!15}-0.44 \\
        $\mathrm{Im}\,a_-$ & \cellcolor{black!3}-0.09 & \cellcolor{black!5}0.17 & \cellcolor{black!5}-0.17 & \cellcolor{black!8}0.24 & \cellcolor{black!7}0.22 & \cellcolor{black!6}-0.17 & \cellcolor{black!6}0.17 & \cellcolor{black!15}-0.44 & \cellcolor{black!35}1.00 & \cellcolor{black!28}-0.82 & \cellcolor{black!8}0.24 & \cellcolor{black!2}0.08 \\
        $\mathrm{Im}\,b_-$ & \cellcolor{black!2}0.07 & \cellcolor{black!9}-0.27 & \cellcolor{black!8}0.26 & \cellcolor{black!11}-0.34 & \cellcolor{black!7}-0.22 & \cellcolor{black!9}0.27 & \cellcolor{black!2}-0.07 & \cellcolor{black!19}0.55 & \cellcolor{black!28}-0.82 & \cellcolor{black!35}1.00 & \cellcolor{black!3}-0.11 & \cellcolor{black!5}-0.15 \\
        $\mathrm{Im}\,a_+$ & \cellcolor{black!0}-0.01 & \cellcolor{black!1}-0.03 & \cellcolor{black!1}0.03 & \cellcolor{black!0}-0.00 & \cellcolor{black!0}0.02 & \cellcolor{black!0}-0.01 & \cellcolor{black!5}-0.16 & \cellcolor{black!9}0.27 & \cellcolor{black!8}0.24 & \cellcolor{black!3}-0.11 & \cellcolor{black!35}1.00 & \cellcolor{black!21}-0.62 \\
        $\mathrm{Im}\,b_+$ & \cellcolor{black!1}-0.06 & \cellcolor{black!7}0.22 & \cellcolor{black!5}-0.17 & \cellcolor{black!3}0.09 & \cellcolor{black!1}0.04 & \cellcolor{black!2}-0.08 & \cellcolor{black!1}-0.03 & \cellcolor{black!15}-0.44 & \cellcolor{black!2}0.08 & \cellcolor{black!5}-0.15 & \cellcolor{black!21}-0.62 & \cellcolor{black!35}1.00 \\
        \bottomrule
    \end{tabular}
    \caption{Correlation matrix of the 12 hadronic parameters obtained from the fit to CP-averaged observables using measurements of LHCb configuration (iv) and sampling the likelihood.}
    \label{tab:correlation_iv_Samples}
\end{table}

\clearpage
\section{Correlation Matrices of the Wilson Coefficients} \label{app:correlations_WC}

In this appendix we collect the correlation matrices of the (reparameterized) Wilson coefficients obtained from the fits to CP-averaged observables and CP asymmetries using measurements of LHCb configuration (iv) in Section~\ref{sec:fits_CPV}.

The correlation matrix from the unconstrained fit is shown in table~\ref{tab:correlation_np_reparam_Samples}, the one from the fit with $\text{Re}\,C_9 = \text{Re}\,C_9^\prime = 0$ is shown in table~\ref{tab:correlation_np_zero_c9_c9p_Samples}, and the one from the fit with hadronic parameters set to the \texttt{flavio} defaults is shown in table~\ref{tab:correlation_np_flavio_Samples}. In each case we show the results that we obtain by sampling the likelihood.
We have checked that using instead the inverse of the Hessian in the best fit point gives consistent results.

\begin{table}[tbh]
    \centering
    \small
    \setlength{\tabcolsep}{1.4pt}
        \begin{tabular}{l|cccc|cccc}
        \toprule
          & $\widetilde{\mathrm{Re}\,C_9}$ & $\widetilde{\mathrm{Re}\,C_9^\prime}$ & $\mathrm{Re}\,C_{10}$ & $\mathrm{Re}\,C_{10}^\prime$ & $\mathrm{Im}\,C_9$ & $\mathrm{Im}\,C_9^\prime$ & $\mathrm{Im}\,C_{10}$ & $\mathrm{Im}\,C_{10}^\prime$ \\
        \midrule
        $\widetilde{\mathrm{Re}\,C_9}$ & \cellcolor{black!35}1.00 & \cellcolor{black!11}-0.34 & \cellcolor{black!0}0.00 & \cellcolor{black!0}-0.03 & \cellcolor{black!8}0.23 & \cellcolor{black!0}-0.01 & \cellcolor{black!0}0.01 & \cellcolor{black!0}0.00 \\
        $\widetilde{\mathrm{Re}\,C_9^\prime}$ & \cellcolor{black!11}-0.34 & \cellcolor{black!35}1.00 & \cellcolor{black!0}-0.01 & \cellcolor{black!0}-0.01 & \cellcolor{black!0}-0.00 & \cellcolor{black!8}0.24 & \cellcolor{black!0}-0.00 & \cellcolor{black!0}0.01 \\
        $\mathrm{Re}\,C_{10}$ & \cellcolor{black!0}0.00 & \cellcolor{black!0}-0.01 & \cellcolor{black!34}1.00 & \cellcolor{black!11}0.32 & \cellcolor{black!2}0.07 & \cellcolor{black!3}0.09 & \cellcolor{black!1}0.05 & \cellcolor{black!2}0.07 \\
        $\mathrm{Re}\,C_{10}^\prime$ & \cellcolor{black!0}-0.03 & \cellcolor{black!0}-0.01 & \cellcolor{black!11}0.32 & \cellcolor{black!35}1.00 & \cellcolor{black!3}0.11 & \cellcolor{black!4}0.12 & \cellcolor{black!3}-0.09 & \cellcolor{black!0}-0.02 \\
        \midrule
        $\mathrm{Im}\,C_9$ & \cellcolor{black!8}0.23 & \cellcolor{black!0}-0.00 & \cellcolor{black!2}0.07 & \cellcolor{black!3}0.11 & \cellcolor{black!34}1.00 & \cellcolor{black!13}0.38 & \cellcolor{black!14}0.40 & \cellcolor{black!16}0.47 \\
        $\mathrm{Im}\,C_9^\prime$ & \cellcolor{black!0}-0.01 & \cellcolor{black!8}0.24 & \cellcolor{black!3}0.09 & \cellcolor{black!4}0.12 & \cellcolor{black!13}0.38 & \cellcolor{black!35}1.00 & \cellcolor{black!7}0.20 & \cellcolor{black!7}0.23 \\
        $\mathrm{Im}\,C_{10}$ & \cellcolor{black!0}0.01 & \cellcolor{black!0}-0.00 & \cellcolor{black!1}0.05 & \cellcolor{black!3}-0.09 & \cellcolor{black!14}0.40 & \cellcolor{black!7}0.20 & \cellcolor{black!35}1.00 & \cellcolor{black!24}0.70 \\
        $\mathrm{Im}\,C_{10}^\prime$ & \cellcolor{black!0}0.00 & \cellcolor{black!0}0.01 & \cellcolor{black!2}0.07 & \cellcolor{black!0}-0.02 & \cellcolor{black!16}0.47 & \cellcolor{black!7}0.23 & \cellcolor{black!24}0.70 & \cellcolor{black!35}1.00 \\
        \bottomrule
    \end{tabular}
    \bigskip
    \caption{Correlation matrix of the reparameterized Wilson coefficients obtained from the unconstrained fit to CP-averaged observables and CP asymmetries, using measurements of LHCb configuration (iv), and sampling the likelihood.}
    \label{tab:correlation_np_reparam_Samples}
\end{table}
\begin{table}[tbh]
    \centering
    \small
    \setlength{\tabcolsep}{1.4pt}
        \begin{tabular}{l|cccc|cccc}
        \toprule
          & $\widetilde{\mathrm{Re}\,C_9}$ & $\widetilde{\mathrm{Re}\,C_9^\prime}$ & $\mathrm{Re}\,C_{10}$ & $\mathrm{Re}\,C_{10}^\prime$ & $\mathrm{Im}\,C_9$ & $\mathrm{Im}\,C_9^\prime$ & $\mathrm{Im}\,C_{10}$ & $\mathrm{Im}\,C_{10}^\prime$ \\
        \midrule
        $\widetilde{\mathrm{Re}\,C_9}$ & -- & -- & -- & -- & -- & -- & -- & -- \\
        $\widetilde{\mathrm{Re}\,C_9^\prime}$ & -- & -- & -- & -- & -- & -- & -- & -- \\
        $\mathrm{Re}\,C_{10}$ & -- & -- & \cellcolor{black!35}1.00 & \cellcolor{black!11}0.33 & \cellcolor{black!2}0.08 & \cellcolor{black!3}0.09 & \cellcolor{black!1}0.03 & \cellcolor{black!2}0.06 \\
        $\mathrm{Re}\,C_{10}^\prime$ & -- & -- & \cellcolor{black!11}0.33 & \cellcolor{black!35}1.00 & \cellcolor{black!4}0.12 & \cellcolor{black!5}0.14 & \cellcolor{black!3}-0.10 & \cellcolor{black!1}-0.04 \\
        \midrule
        $\mathrm{Im}\,C_9$ & -- & -- & \cellcolor{black!2}0.08 & \cellcolor{black!4}0.12 & \cellcolor{black!35}1.00 & \cellcolor{black!13}0.39 & \cellcolor{black!14}0.42 & \cellcolor{black!17}0.49 \\
        $\mathrm{Im}\,C_9^\prime$ & -- & -- & \cellcolor{black!3}0.09 & \cellcolor{black!5}0.14 & \cellcolor{black!13}0.39 & \cellcolor{black!35}1.00 & \cellcolor{black!6}0.20 & \cellcolor{black!7}0.22 \\
        $\mathrm{Im}\,C_{10}$ & -- & -- & \cellcolor{black!1}0.03 & \cellcolor{black!3}-0.10 & \cellcolor{black!14}0.42 & \cellcolor{black!6}0.20 & \cellcolor{black!35}1.00 & \cellcolor{black!24}0.71 \\
        $\mathrm{Im}\,C_{10}^\prime$ & -- & -- & \cellcolor{black!2}0.06 & \cellcolor{black!1}-0.04 & \cellcolor{black!17}0.49 & \cellcolor{black!7}0.22 & \cellcolor{black!24}0.71 & \cellcolor{black!34}1.00 \\
        \bottomrule
    \end{tabular}
    \bigskip
    \caption{Correlation matrix of the Wilson coefficients obtained from the fit to CP-averaged observables and CP asymmetries, with $\text{Re}\,C_9=\text{Re}\,C_9^\prime=0$, using measurements of LHCb configuration (iv),  and sampling the likelihood.}
    \label{tab:correlation_np_zero_c9_c9p_Samples}
\end{table}
\begin{table}[tbh]
    \centering
    \small
    \setlength{\tabcolsep}{1.4pt}
        \begin{tabular}{l|cccc|cccc}
        \toprule
          & $\mathrm{Re}\,C_9$ & $\mathrm{Re}\,C_9^\prime$ & $\mathrm{Re}\,C_{10}$ & $\mathrm{Re}\,C_{10}^\prime$ & $\mathrm{Im}\,C_9$ & $\mathrm{Im}\,C_9^\prime$ & $\mathrm{Im}\,C_{10}$ & $\mathrm{Im}\,C_{10}^\prime$ \\
        \midrule
        $\mathrm{Re}\,C_9$ & \cellcolor{black!35}1.00 & \cellcolor{black!7}0.20 & \cellcolor{black!3}-0.11 & \cellcolor{black!8}0.25 & \cellcolor{black!0}-0.00 & \cellcolor{black!9}0.26 & \cellcolor{black!1}0.03 & \cellcolor{black!0}0.01 \\
        $\mathrm{Re}\,C_9^\prime$ & \cellcolor{black!7}0.20 & \cellcolor{black!35}1.00 & \cellcolor{black!3}0.09 & \cellcolor{black!7}0.23 & \cellcolor{black!1}-0.05 & \cellcolor{black!3}0.09 & \cellcolor{black!1}-0.05 & \cellcolor{black!2}-0.06 \\
        $\mathrm{Re}\,C_{10}$ & \cellcolor{black!3}-0.11 & \cellcolor{black!3}0.09 & \cellcolor{black!34}1.00 & \cellcolor{black!12}0.36 & \cellcolor{black!3}0.11 & \cellcolor{black!4}0.13 & \cellcolor{black!2}0.07 & \cellcolor{black!3}0.09 \\
        $\mathrm{Re}\,C_{10}^\prime$ & \cellcolor{black!8}0.25 & \cellcolor{black!7}0.23 & \cellcolor{black!12}0.36 & \cellcolor{black!35}1.00 & \cellcolor{black!8}0.25 & \cellcolor{black!5}0.16 & \cellcolor{black!2}0.07 & \cellcolor{black!3}0.11 \\
        \midrule
        $\mathrm{Im}\,C_9$ & \cellcolor{black!0}-0.00 & \cellcolor{black!1}-0.05 & \cellcolor{black!3}0.11 & \cellcolor{black!8}0.25 & \cellcolor{black!35}1.00 & \cellcolor{black!5}0.16 & \cellcolor{black!19}0.54 & \cellcolor{black!20}0.59 \\
        $\mathrm{Im}\,C_9^\prime$ & \cellcolor{black!9}0.26 & \cellcolor{black!3}0.09 & \cellcolor{black!4}0.13 & \cellcolor{black!5}0.16 & \cellcolor{black!5}0.16 & \cellcolor{black!35}1.00 & \cellcolor{black!8}0.25 & \cellcolor{black!9}0.27 \\
        $\mathrm{Im}\,C_{10}$ & \cellcolor{black!1}0.03 & \cellcolor{black!1}-0.05 & \cellcolor{black!2}0.07 & \cellcolor{black!2}0.07 & \cellcolor{black!19}0.54 & \cellcolor{black!8}0.25 & \cellcolor{black!34}1.00 & \cellcolor{black!27}0.79 \\
        $\mathrm{Im}\,C_{10}^\prime$ & \cellcolor{black!0}0.01 & \cellcolor{black!2}-0.06 & \cellcolor{black!3}0.09 & \cellcolor{black!3}0.11 & \cellcolor{black!20}0.59 & \cellcolor{black!9}0.27 & \cellcolor{black!27}0.79 & \cellcolor{black!35}1.00 \\
        \bottomrule
    \end{tabular}
    \bigskip
    \caption{Correlation matrix of the Wilson coefficients obtained from the fit to CP-averaged observables and CP asymmetries, with \texttt{flavio} default hadronic parameters, using measurements of LHCb configuration (iv), and sampling the likelihood.}
    \label{tab:correlation_np_flavio_Samples}
\end{table}
\end{appendix}

\clearpage
\bibliographystyle{JHEP}
\bibliography{bibliography.bib}

\end{document}